\documentclass[twoside]{article}
\usepackage{ecj,palatino,latexsym,natbib}

\usepackage{epsfig,wrapfig,boxedminipage,amssymb,amsmath,amscd}
\usepackage{capt-of}\usepackage{times} 
\usepackage{epstopdf}
\usepackage[T1]{fontenc}
\usepackage[usenames]{color}
\usepackage[official,right]{eurosym}
\usepackage{hyperref}

\def\reals{\mathbb{R}}

\def\comp{\raise 1pt \hbox{$\scriptstyle\circ$}}

\def\argmin{\mathop{\rm argmin}\limits}
\def\argmax{\mathop{\rm argmax}\limits}

\def\upto{{\raise 1pt \hbox{$\scriptstyle \,\nearrow\,$}}}
\def\downto{{\raise 1pt \hbox{$\scriptstyle \,\searrow\,$}}}

\def\tos{\rightrightarrows}

\newtheorem{theorem}{Theorem}

\newtheorem{definition}[theorem]{Definition}

\begin{document}

\newcommand{\boldx}{\mbox{$\mathbf{x}$}}
\newcommand{\boldy}{\mbox{$\mathbf{y}$}}
\newcommand{\boldf}{\mbox{$\mathbf{f}$}}
\newcommand{\boldz}{\mbox{$\mathbf{z}$}}
\newcommand{\boldF}{\mbox{$\mathbf{F}$}}
\newcommand{\boldG}{\mbox{$\mathbf{G}$}}
\newcommand{\boldg}{\mbox{$\mathbf{g}$}}
\newcommand{\boldh}{\mbox{$\mathbf{h}$}}
\newcommand{\boldH}{\mbox{$\mathbf{H}$}}
\newcommand{\boldzero}{\mbox{$\mathbf{0}$}}
\newcommand{\Rbb}{\mbox{$\mathbb R$}}

\newtheorem{defn}{Definition}

\title{\bf Test Problem Construction for Single-Objective Bilevel Optimization}  

\author{\name{\bf Ankur Sinha} \hfill \addr{ankur.sinha@aalto.fi}\\ 
        \addr{Department of Information and Service Economy, Aalto University School of Business, 
        Helsinki, 00076 Aalto, Finland}
\AND
       \name{\bf Pekka Malo} \hfill \addr{pekka.malo@aalto.fi}\\
        \addr{Department of Information and Service Economy, Aalto University School of Business, 
        Helsinki, 00076 Aalto, Finland}
\AND
       \name{\bf Kalyanmoy Deb\thanks{\hspace{1mm} Also Visiting Professor at Aalto University School of Business, Helsinki, 00076 Aalto, Finland}} \hfill \addr{kdeb@egr.msu.edu}\\
        \addr{Department of Electrical and Computer Engineering, Michigan State University, 
        East Lansing, MI 48823, USA}
}

\maketitle


\begin{abstract}
In this paper, we propose a procedure for designing controlled test problems for single-objective bilevel optimization. The construction procedure is flexible and allows its user to control the different complexities that are to be included in the test problems independently of each other. In addition to properties that control the difficulty in convergence, the procedure also allows the user to introduce difficulties caused by interaction of the two levels. As a companion to the test problem construction framework, the paper presents a standard test suite of twelve problems, which includes eight unconstrained and four constrained problems. Most of the problems are scalable in terms of variables and constraints. To provide baseline results, we have solved the proposed test problems using a nested bilevel evolutionary algorithm. The results can be used for comparison, while evaluating the performance of any other bilevel optimization algorithm. The codes related to the paper may be accessed from the website \url{http://bilevel.org}.
\end{abstract}


\section{Introduction}
Bilevel optimization constitutes a challenging class of optimization problems, where one optimization task is nested within the other. A large number of studies have been conducted in the field of bilevel programming \citep{colson,vicente-review,dempe-dutta,my-ecj10}, and on its practical applications \citep{dempe02}. Classical approaches commonly used to handle bilevel problems include the Karush-Kuhn-Tucker approach \citep{bianco-kkt,bilevel-KKT1}, branch-and-bound techniques \citep{bard82} and the use of penalty functions \citep{aiyoshi81}. Despite a significant progress made in classical optimization towards solving bilevel optimization problems, most of these approaches are rendered inapplicable for bilevel problems with higher levels of complexity. Over the last two decades, technological advances and availability of enormous computing resources have given rise to heuristic approaches for solving difficult optimization problems. Heuristics such as evolutionary algorithms are recognized as potent tools for handling challenging classes of optimization problems. A number of studies have been performed towards using evolutionary algorithms \citep{yin-bilevel,GA_Wang,my-ecj10} for solving bilevel problems. However, the research on evolutionary algorithms for bilevel problems is still in nascent stage, and significant improvement in the existing approaches is required. Most of the heuristic approaches lack a finite time convergence proof for optimization problems. Therefore, it is a common practice among researchers to demonstrate the convergence of their algorithms on a test bed constituting problems with various complexities. To the best of our knowledge, there does not exist a systematic framework for constructing single-objective bilevel test problems with controlled difficulties. Test problems, which offer various difficulties found in practical application problems, are often required during the construction and evaluation of algorithms.

Past studies \citep{mitsos06} on bilevel optimization have introduced a number of simple test problems. However, the levels of difficulty cannot be controlled in these test problems. In most of the studies, the problems are either linear \citep{moshirvaziri96}, or quadratic \citep{calamai92,calamai94}, or non-scalable with fixed number of decision variables.
Application problems in transportation (network design, optimal pricing) \citep{migdalas95,constantin95,brotcorne01}, economics (Stackelberg games, principal-agent problem, taxation, policy decisions) \citep{fudenberg93,stackelbergWang01,my-caor13,my-cec13}, management (network facility location, coordination of multi-divisional firms) \citep{sun08,bard83}, engineering (optimal design, optimal chemical equilibria) \citep{kirjnerneto98,smith82} have also been used to demonstrate the efficiency of algorithms. For most real-world problems, the true optimal solution is unknown. Therefore, it is hard to identify, whether a particular solution obtained using an existing approach is close to the optima. Under these uncertainties, it is not possible to systematically evaluate solution procedures on practical problems. These drawbacks pose hurdles in algorithm development, as the performance of the algorithms cannot be evaluated on various difficulty frontiers. A test-suite with controllable level of difficulties helps in understanding the bilevel algorithms better. It gives information on what properties of bilevel problems are handled efficiently by the algorithm and what are not. An algorithm which performs well on the test problem by effectively tackling most of the challenges offered by the test-suite is expected to perform good on other simpler problems as well. Therefore, controlled test problems are necessary to advance the research on bilevel optimization using evolutionary algorithms.

In this paper, we identify the challenges that are commonly encountered in bilevel optimization problems. Based on these findings, we propose a procedure for constructing test problems that mimic these difficulties in a controllable manner. Using the construction procedure, we propose a collection of bilevel test problems that are scalable in terms of variables and constraints. The proposed scheme allows the user to control the difficulties at the two levels independently of each other. At the same time, it also allows the control of the extent of difficulty arising due to interaction of the two levels. To make algorithm evaluation easier, the problems generated using the framework are such that the optimal solution of the bilevel problem is known. Moreover, the induced set of the bilevel problem is known as a function of the upper level variables. Such information helps the algorithm developers to debug their procedures during the development phase, and also allows to evaluate the convergence properties of the approach. 

The paper is organized as follows. In the next section, we explain the structure of a general bilevel optimization problem and introduce the notation used in the paper. Section~\ref{sec:framework} presents our framework for constructing scalable test problems for bilevel programming. Thereafter, following the guidelines of the construction procedure, we suggest a set of twelve scalable test problems in Section~\ref{sec:testproblems}. To create a benchmark for evaluating different solution algorithms, the problems are solved using a simple nested bilevel evolutionary algorithm which is a nested scheme described in Section~\ref{sec:nested-algorithm}. The results for the baseline algorithm are discussed in Section~\ref{sec:results}.

\section{Description of a Bilevel Problem}
A bilevel optimization problem involves two levels of optimization tasks, where one level is nested within the other. The outer optimization task is usually called upper level optimization task, and the inner optimization task is called lower level optimization task. The hierarchical structure of the problem requires that only the optimal solutions of the inner optimization task are acceptable as feasible members for the outer optimization task. The problem contains two types of variables; namely the upper level variables $\boldx_u$, and the lower level variables $\boldx_l$. The lower level is optimized with respect to the lower level variables $\boldx_l$, and the upper level variables $\boldx_u$ act as parameters. An optimal lower level vector and the corresponding upper level vector $\boldx_u$ constitute a feasible upper level solution, provided the upper level constraints are also satisfied. The upper level problem involves  all variables $\boldx=(\boldx_u,\boldx_l)$, and the optimization is to be performed with respect to both $\boldx_u$ and $\boldx_l$. In the following, we provide two equivalent formulations for a general bilevel optimization problem with one objective at both levels:

\vskip 0.1cm
\begin{definition}[Bilevel Optimization Problem (BLOP)]
Let $X=X_U\times X_L$ denote the product of the upper-level decision space  $X_U$ and the lower-level decision space $X_L$, i.e. $\boldx=(\boldx_u,\boldx_l)\in X$, if $\boldx_u\in X_U$ and $\boldx_l\in X_L$. For upper-level objective function $F:X\to \reals$ and lower-level objective function $f:X\to \reals$, a general bilevel optimization problem is given by
\begin{equation}
\begin{array}{rl}
\underset{{\boldx\in X}}{\mbox{Min}} & F(\boldx), \\
\mbox{s.t.} & 
\boldx_l \in \underset{\boldx_l\in X_L}{\argmin}\left\lbrace 
 f(\boldx) \hspace{1mm}\big|\hspace{1mm} g_i(\boldx) \geq 0, i\in I \right\rbrace, \\
& G_j(\boldx)\geq 0, j\in J. 
\end{array} 
\label{eq:bilevel_multi_obj1}
\end{equation}
where the functions $g_i:X\to\reals$, $i\in I$, represent lower-level constraints and $G_j:X\to\reals$, $j\in J$, is the collection of upper-level constraints.
\end{definition}
\vskip 0.2cm

In the above formulation, a vector $\boldx^{(0)}=(\boldx_{u}^{(0)},\boldx_{l}^{(0)})$ is considered feasible at the upper level, if it satisfies all the upper level constraints, and vector $\boldx_{l}^{(0)}$ is optimal at the lower level for the given $\boldx_{u}^{(0)}$.
We observe in this formulation that the lower-level problem is a parameterized constraint to the upper-level problem. An equivalent formulation of the bilevel optimization problem is obtained by replacing the lower-level optimization problem with a set value function which maps the given upper-level decision vector to the corresponding set of optimal lower-level solutions. In the domain of Stackelberg games, such mapping is referred as the rational reaction of the follower to the leader's choice $\boldx_u$.

\vskip 0.2cm
\begin{definition}[Alternative definition of Bilevel Problem] Let set-valued function $\Psi:X_U\tos X_L$, denote the optimal-solution set mapping of the lower level problem, i.e.
$$
\Psi(\boldx_u)=\underset{\boldx_l\in X_L}{\argmin}\left\lbrace f(\boldx) \hspace{1mm}\big|\hspace{1mm} g_i(\boldx) \geq 0, i\in I \right\rbrace.
$$ 
A general bilevel optimization problem (BLOP) is then given by
\begin{equation}
\begin{array}{rl}
\underset{{\boldx\in X}}{\mbox{Min}} & F(\boldx), \\
\mbox{s.t.} & 
\boldx_l \in \Psi(\boldx_u), \\
& G_j(\boldx)\geq 0, j\in J.
\end{array} 
\label{eq:bilevel_multi_obj2}
\end{equation}
where the function $\Psi$ may be a single-vector valued or a multi-vector valued function depending on whether the lower level function has multiple global optimal solutions or not.
\end{definition}
\vskip 0.2cm

\begin{figure*}
\begin{minipage}[t]{0.49\linewidth}
\begin{center}
\epsfig{file=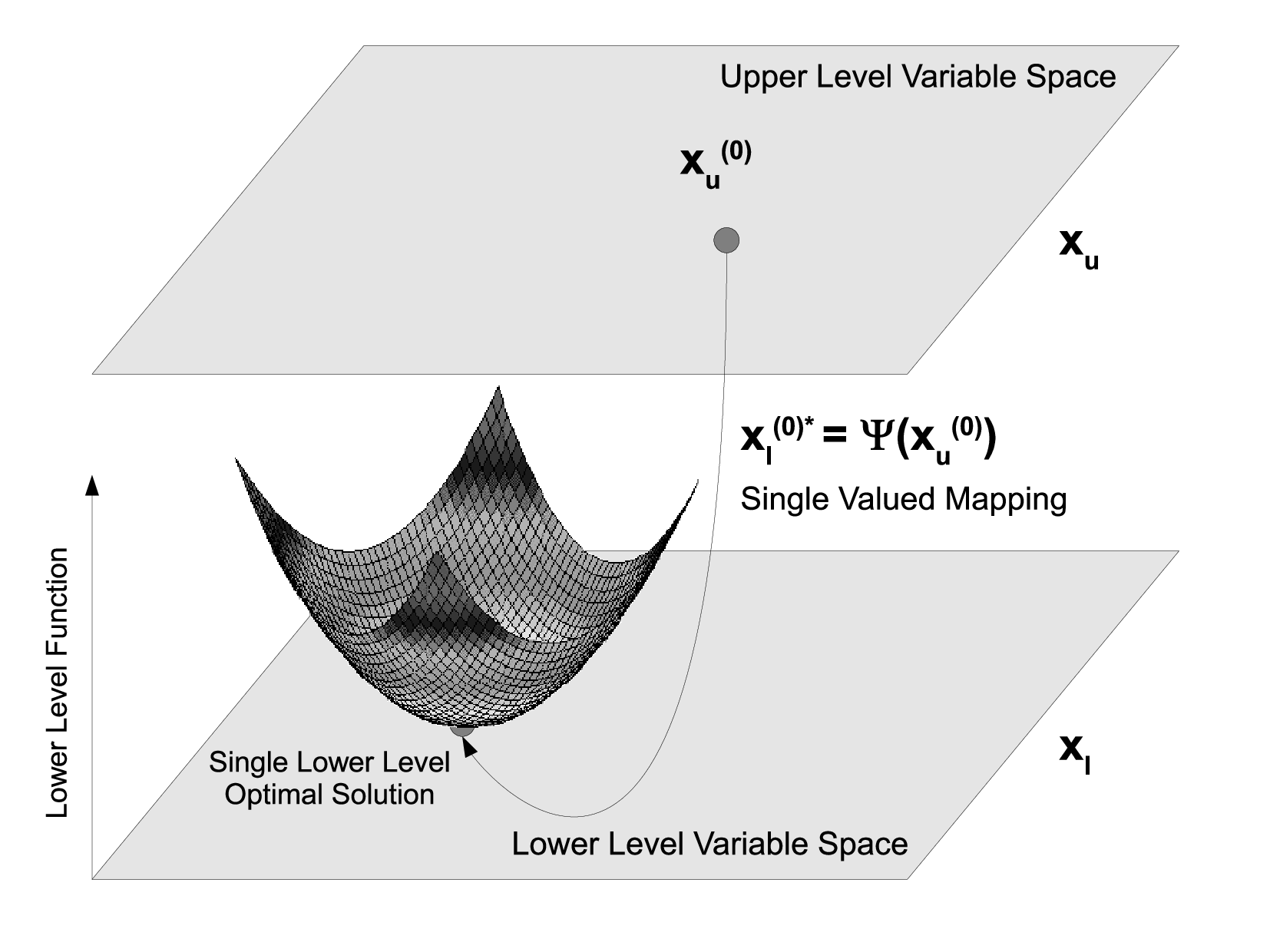,width=\linewidth}
\end{center}
\caption{Relationship between upper and lower level variables in case of a single-vector valued mapping. For simplicity the lower level function has the shape of a paraboloid.}
\label{fig:figure1_par}
\end{minipage}\hfill
\begin{minipage}[t]{0.49\linewidth}
\begin{center}
\epsfig{file=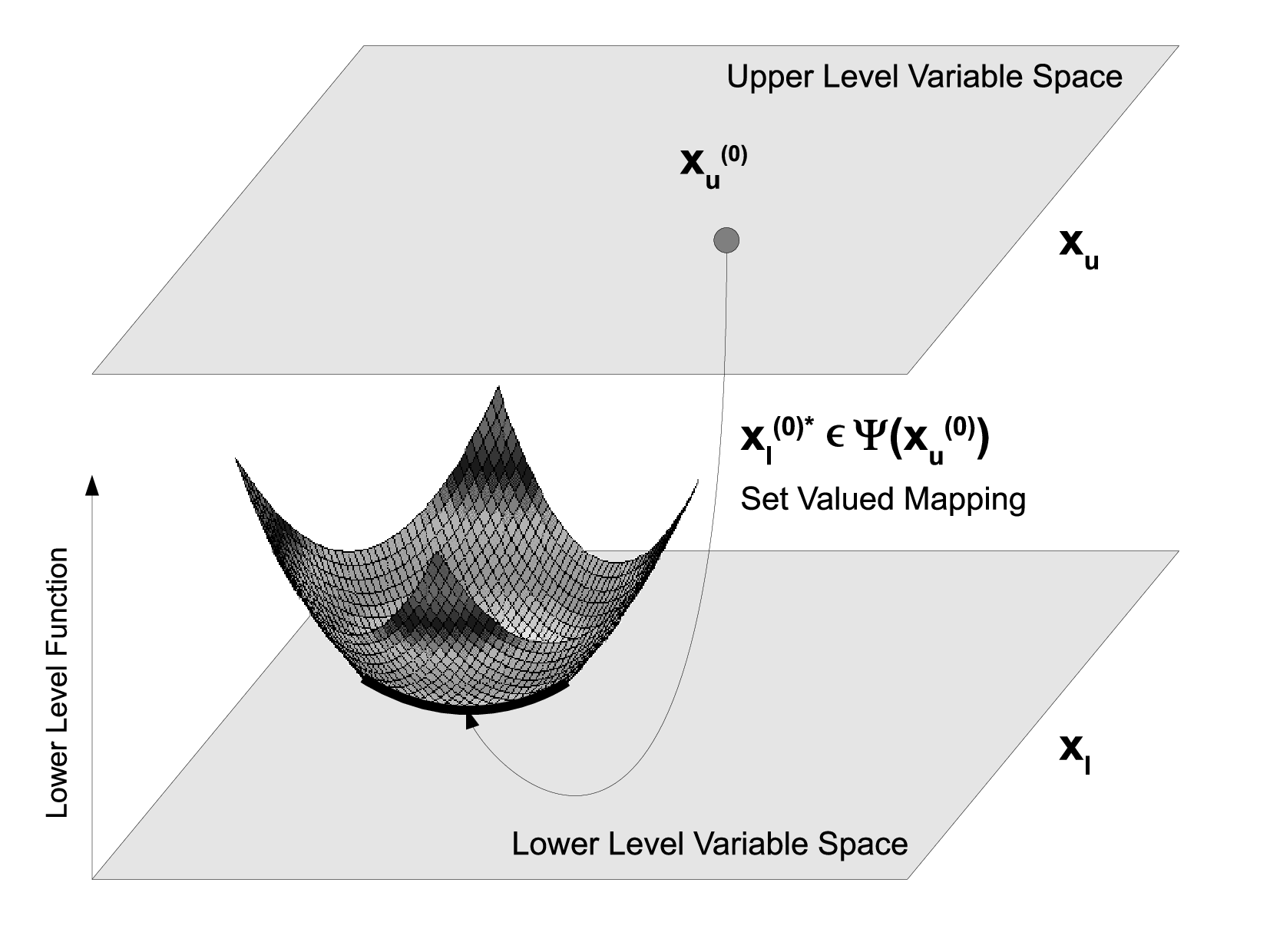,width=\linewidth} 
\end{center}
\caption{Relationship between upper and lower level variables in case of a multi-vector valued mapping. The lower level function is shown in the shape of a paraboloid with the bottom sliced with a plane.}
\label{fig:figure2_par}
\end{minipage}
\end{figure*}

In the test problem construction procedure, the $\Psi$ function provides a convenient description of the relationship between the upper and lower level problems. Figures~\ref{fig:figure1_par} and \ref{fig:figure2_par} illustrate two scenarios, where $\Psi$ can be a single vector valued or a multi-vector valued function respectively. In Figure~\ref{fig:figure1_par}, the lower level problem is shown to be a paraboloid with a single minimum function value corresponding to the set of upper level variables $\boldx_u$. Figure~\ref{fig:figure2_par} represents a scenario where the lower level function is a paraboloid sliced from the bottom with a horizontal plane. This leads to multiple minimum values for the lower level problem, and therefore, multiple lower level solutions correspond to the set of upper level variables $\boldx_u$.

Before discussing the test problem construction framework, we provide further insights into bilevel programming through a simple real-world problem \citep{my-caor13,my-ifac12}. The problem is chosen from the domain of game theory, where there are two entities in Stackelberg competition with each other. The upper level entity is a leader firm and the lower level entity is a follower firm. The leader and the follower firms compete with each other in order to maximize their profits $\Pi_{l}$ and $\Pi_{f}$ respectively. The leader makes the first move and therefore has the first mover's advantage. For any given action of the leader firm, the follower firm reacts optimally. 
With complete knowledge about the follower firm, the leader firm solves the following bilevel optimization problem in order to determine the Stackelberg optimum.
\begin{eqnarray} \label{eq:simple}
	\max_{q_l,q_f,Q} \quad&& \Pi_l = P(Q) q_l - C(q_l) \\
	\mbox{s.t.} \quad&& q_f \in \argmax_{q_f} \lbrace {\Pi_f = P(Q) q_f - C(q_f)} \rbrace , \\
	&& q_l + q_f \geq Q , \label{eq:demand}\\
	&& q_l, q_f, Q \geq 0 ,
\end{eqnarray}
where $Q$ is the quantity demanded, $P(q_l,q_f)$ is the price of the goods sold, and $C(\cdot)$ is the cost of production of the respective firm. The variables in this model are the production levels of each firm $q_l$, $q_f$ and demand $Q$. The leader sets its production level first, and then the follower chooses its production level based on the leader's decision. This simple model assumes homogeneity of the products manufactured by the firms. Additionally, constraint (\ref{eq:demand}) ensures that all demand is satisfied.
By assuming that the firms produce and sell homogeneous goods, we specify a single linear price function for both firms as an inverse demand function of the form
\begin{equation} \label{eq:simpleprice}
	P(Q) = \alpha - \beta Q ,
\end{equation}
where $\alpha, \beta > 0$ are constants. Since costs often tend to increase with the amount of production, we assume convex quadratic cost functions for both firms to be of the form
\begin{align} \label{eq:simplecosts}
	C(q_l) & = \delta_l q_l^2 + \gamma_l q_l + c_l, \\
	C(q_f) & = \delta_f q_f^2 + \gamma_f q_f + c_f,
\end{align}
where $c_i$ denote the fixed costs of the respective firm, and $\delta_i$ and $\gamma_i$ are positive constants. It is possible to solve this bilevel problem analytically. The optimal strategies of the leader and follower, $(q_l^*,q_f^*)$, in this simple linear-quadratic model can be found by using simple differentiation. For brevity, we avoid the steps and directly provide the analytical optimum for the problem.
\begin{equation}
	q_l^* = \frac{2(\beta + \delta_f)(\alpha - \gamma_l) - \beta(\alpha - \gamma_f)}{4(\beta + \delta_f)(\beta + \delta_l) - 2\beta^2}. \label{eq:optlead}
\end{equation}
\begin{equation}
	q_f^* = \frac{\alpha - \gamma_f}{2(\beta + \delta_f)} - \frac{\beta(\alpha - \gamma_l)-{\displaystyle \frac{\beta^2(\alpha - \gamma_f)}{2(\beta + \delta_f)}}}{4(\beta + \delta_f)(\beta + \delta_l) - 2\beta^2}. \label{eq:optfol}
\end{equation}
Equations (\ref{eq:optlead}) and (\ref{eq:optfol}) are the strategies of the leader and follower at Stackelberg equilibrium. These depend only on the constant parameters of the model. Given these values, the leader will choose the production level given by Equation (\ref{eq:optlead}), and the follower will react optimally by choosing its production level using Equation (\ref{eq:optfol}). At the optimum, constraint (\ref{eq:demand}) holds as a strict equality, which provides us the optimal demand $Q^*$. In the presence of linear and quadratic functions, it is possible to solve the model analytically. However, as soon as the functions get complicated, it becomes difficult to find the optimum using analytical or numerical approaches. Next, we provide a test problem construction framework that allows us to create scalable bilevel test problems with a variety of difficulties commonly encountered in bilevel optimization.

\section{Test Problem Construction Procedure}\label{sec:framework}
The presence of an additional optimization task within the constraints of the upper level optimization task leads to a significant increase in complexity, as compared to any single level optimization problem. In this section, we describe various kinds of complexities, which a bilevel optimization problem can offer, and provide a test problem construction procedure that can induce these difficulties in a controllable manner. In order to create realistic test problems, the construction procedure should be able to control the scale of difficulties at both levels independently and collectively, such that the performance of algorithms in handling the two levels is evaluated. The test problems created using the construction procedure are expected to be scalable in terms of number of decision variables and constraints, such that the performance of the algorithms can be evaluated against increasing number of variables and constraints. The construction procedure should be able to generate test problems with the following properties:

\vspace{2mm}

\hspace{-7mm}{\bf Necessary Properties:}
\begin{enumerate}
\item The optimal solution of the bilevel optimization should be known.
\item Clear identification of a relationship between the lower level optimal solutions and the upper level variables.
\end{enumerate}

\hspace{-7mm}{\bf Properties for inducing difficulties:}
\begin{enumerate}
\item Controlled difficulty in convergence at upper and lower levels.
\item Controlled difficulty caused by interaction of the two levels.
\item Multiple global solutions at the lower level for a given set of upper level variables.
\item Possibility to have either conflict or cooperation between the two levels.
\item Scalability to any number of decision variables at upper and lower levels.
\item Constraints (preferably scalable) at upper and lower levels.
\end{enumerate}

Next, we provide the bilevel test problem construction procedure, which is able to induce most of the difficulties suggested above.

\subsection{Objective functions in the test-problem framework}

To create a tractable framework for test-problem construction, we split the upper and lower level functions into three components. Each of the components is specialized for induction of certain kinds of difficulties into the bilevel problem. The functions are determined based on the required complexities at upper and lower levels independently, and also by the required complexities because of the interaction of the two levels. In this setting, a generic bilevel test problem can be written as follows:

\begin{equation}
\begin{array}{l}
F(\boldx_u,\boldx_l) = F_1(\boldx_{u1}) + F_2(\boldx_{l1}) + F_3(\boldx_{u2},\boldx_{l2}) \\
f(\boldx_u,\boldx_l) = f_1(\boldx_{u1}, \boldx_{u2}) + f_2(\boldx_{l1}) + f_3(\boldx_{u2},\boldx_{l2})\\

\mbox{where}\\
     \quad \quad \boldx_u = (\boldx_{u1}, \boldx_{u2}) \quad \mbox{and} \quad \boldx_l = (\boldx_{l1}, \boldx_{l2})
\end{array}
\end{equation}

\begin{table*}
\caption{Overview of test-problem framework components}\label{tab:framework}
\begin{minipage}{1.0\linewidth}
\begin{footnotesize}
\begin{center}
\begin{tabular}{|c|c|c|c|}
\multicolumn{4}{c}{Panel A: Decomposition of decision variables}\\
\hline
\multicolumn{2}{|c|}{Upper-level variables} & \multicolumn{2}{|c|}{Lower-level variables} \\
\hline
Vector & Purpose & Vector & Purpose \\
\hline\hline
$\boldx_{u1}$ & Complexity on upper-level & $\boldx_{l1}$ & Complexity on lower-level \\
$\boldx_{u2}$ & Interaction with lower-level & $\boldx_{l2}$ & Interaction with upper-level \\
\hline

\multicolumn{4}{c}{}\\

\multicolumn{4}{c}{Panel B: Decomposition of objective functions}\\
\hline
\multicolumn{2}{|c|}{Upper-level objective function} & \multicolumn{2}{|c|}{Lower-level objective function} \\
\hline
Component & Purpose & Component & Purpose \\
\hline\hline
 $F_1(\boldx_{u1})$& Difficulty in convergence & $f_1(\boldx_{u1}, \boldx_{u2})$ & Functional dependence \\
 $F_2(\boldx_{l1})$& Conflict / co-operation & $f_2(\boldx_{l1})$ & Difficulty in convergence\\
 $F_3(\boldx_{u2},\boldx_{l2})$& Difficulty in interaction & $f_3(\boldx_{u2},\boldx_{l2})$ & Difficulty in interaction \\
\hline
\end{tabular}
\end{center}
\end{footnotesize}
\end{minipage}
\end{table*}

In the above equations, each of the levels contains three terms. A summary on the roles of different terms is provided in Table~\ref{tab:framework}. The upper level and lower level variables have been broken into two smaller vectors (see Panel A in Table~\ref{tab:framework}). The vectors $\boldx_{u1}$ and $\boldx_{l1}$ are used to induce complexities at the upper and lower levels independently. The vectors $\boldx_{u2}$ and $\boldx_{l2}$ are responsible to induce complexities because of interaction. In a similar fashion, we decompose the upper and lower level functions such that each of the components is specialized for a certain purpose only (see Panel B in Table~\ref{tab:framework}). At the upper level, the term $F_1(\boldx_{u1})$ is responsible for inducing difficulty in convergence solely at the upper level. Similarly, at the lower level, the term $f_2(\boldx_{l1})$ is responsible for inducing difficulty in convergence solely at the lower level. The term $F_2(\boldx_{l1})$ decides if there is a conflict or a cooperation between the upper and lower levels. The terms $F_3(\boldx_{l2}, \boldx_{u2})$ and $f_3(\boldx_{l2}, \boldx_{u2})$ are interaction terms which can be used to induce difficulties because of interaction at the two levels. Term $F_3(\boldx_{l2}, \boldx_{u2})$ may also induce a cooperation or a conflict. Finally, $f_1(\boldx_{u1}, \boldx_{u2})$ is a fixed term for the lower level optimization problem and does not induce any convergence difficulties. It is used along with the lower level interaction term to create a functional dependence between lower level optimal solution(s) and the upper level variables. The difficulties related to constraints are handled separately.

\subsubsection{Controlled difficulty in convergence}

The test-problem framework allows introduction of difficulties in terms of convergence at both levels of a bilevel optimization problem while retaining sufficient control. To demonstrate this, let us consider the structure of the lower level minimization problem. 
For a given $\boldx_u=(\boldx_{u1},\boldx_{u2})$, the lower level minimization problem is written as,
$$
\underset{(\boldx_{l1},\boldx_{l2})}{\mbox{Min }} f(\boldx_u,\boldx_l)= f_1(\boldx_{u1}, \boldx_{u2}) + f_2(\boldx_{l1}) + f_3(\boldx_{u2},\boldx_{l2}),
$$
where the upper level variables $(\boldx_{u1},\boldx_{u2})$ act as parameters for the optimization problem. The corresponding optimal-set mapping is given by,
$$
\Psi(\boldx_u)=\argmin\{f_2(\boldx_{l1})+f_3(\boldx_{u2},\boldx_{l2}):\boldx_l\in X_L\},
$$
where $f_1$ does not appear due to its independence from $\boldx_l$. Since all of the terms are independent of each other, we note that the optimal value of the function $f$ can be recovered by optimizing the functions $f_2$ and $f_3$ individually. Function $f_2$ contains only lower level variables $\boldx_{l1}$, which do not interact with upper level variables. It introduces convergence difficulties at the lower level without affecting the upper level optimization task. Function $f_3$ contains both lower level variables $\boldx_{l2}$, and upper level variables $\boldx_{u2}$. The optimal value of this function depends on $\boldx_{u2}$.

The following example shows that the calibration of the desired difficulty level for the lower level problem boils down to the choice of functions $f_2$ and $f_3$ such that their optima are known.

\vskip 0.2cm
\textit{Example 1:}\label{ex:simple-lower}
To create a simple lower level function, let the dimension of the variable sets be as follows: $dim(\boldx_{u1}) = U1$, $dim(\boldx_{u2}) = U2$, $dim(\boldx_{l1}) = L1$ and $dim(\boldx_{l2}) = L2$. Consider a special case where $L2 = U2$, then the three functions could be defined as follows,
\begin{equation*}
\begin{array}{l}
f_1(\boldx_{u1}, \boldx_{u2}) = \sum_{i=1}^{U1} (x_{u1}^{i})^2 + \sum_{i=1}^{U2} (x_{u2}^{i})^2,\\
f_2(\boldx_{l1}) =  \sum_{i=1}^{L1} (x_{l1}^{i})^2,\\
f_3(\boldx_{u2},\boldx_{l2}) = \sum_{i=1}^{U2} (x_{u2}^{i} - x_{l2}^{i})^2,\\
\end{array}
\end{equation*}
where $f_1$ affects only the value of the function without inducing any convergence difficulties. The corresponding optimal set mapping $\Psi$ is reduced to an ordinary vector valued function,
$$
\Psi(\boldx_u)=\{(\boldx_{l1},\boldx_{l2}):\boldx_{l1}=\boldzero, \boldx_{l2}=\boldx_{u2}\}.
$$

As discussed above, other functions can be chosen with desired complexities to induce difficulties at the lower level and come up with a variety of lower level functions. Similarly, $F_1$ is a function of $\boldx_{u1}$, which does not interact with any lower level variables. It causes convergence difficulties at the upper level without introducing any other form of complexity in the bilevel problem.

\subsubsection{Controlled difficulty in interaction}
Next, we consider difficulties due to interaction between the upper and lower level optimization tasks. The upper level optimization task is defined as a minimization problem over the graph of the optimal solution set mapping $\Psi$, i.e.,
$$
\mbox{Min }\{F(\boldx_u,\boldx_l):\boldx_l\in\Psi(\boldx_u),\boldx_u\in X_U\},
$$ 
where the objective function
$F(\boldx_{u},\boldx_{l}) = F_1(\boldx_{u1}) + F_2(\boldx_{l1}) + F_3(\boldx_{u2},\boldx_{l2})$
is a sum of three independent terms. Our primary interest is on the last two terms $F_2(\boldx_{l1})$ and $F_3(\boldx_{u2},\boldx_{l2})$, which determine the type of interaction there is going to be between the optimization problems. This can be done in two different ways, depending on whether a cooperation or a conflict is desired between the upper and lower level problems.  

\vskip 0.2cm
\begin{definition}[Co-operative bilevel test-problem]
A bilevel optimization problem is said to be co-operative, if in the vicinity of $\boldx_{l}^{\ast}$ for a particular $\boldx_u$, an improvement in the lower level function value leads to an improvement in the upper level function value.  Within our test problem framework, the independence of terms in the upper level objective function $F$ implies that the co-operative condition is satisfied when for any upper level decision $\boldx_u$ the corresponding lower level decision $\boldx_l=(\boldx_{l1},\boldx_{l2})$ is such that $\boldx_{l1}\in\argmin\{F_2(\boldx_{l1}):\boldx_l\in\Psi(\boldx_u)\}$ and $\boldx_{l2}\in\argmin\{F_3(\boldx_{u2},\boldx_{l2}):\boldx_l\in\Psi(\boldx_u)\}$.
\end{definition}
\vskip 0.2cm

\vskip 0.2cm
\begin{definition}[Conflicting bilevel test-problem]
A bilevel optimization problem is said to be conflicting, if in the vicinity of $\boldx_{l}^{\ast}$ for a particular $\boldx_u$, an improvement in the lower-level function value leads to an adverse effect on the upper level function value. In our framework, a conflicting test problem is obtained when for any upper level decision $\boldx_u$ the corresponding lower level decision $\boldx_l=(\boldx_{l1},\boldx_{l2})$ is such that $\boldx_{l1}\in\argmax\{F_2(\boldx_{l1}):\boldx_l\in\Psi(\boldx_u)\}$ and $\boldx_{l2}\in\argmax\{F_3(\boldx_{u2},\boldx_{l2}):\boldx_l\in\Psi(\boldx_u)\}$.
\end{definition}
\vskip 0.2cm

In the above general form, the functions $f_2$ and $f_3$ may have multiple optimal solutions for any given upper level decision $\boldx_u$. However, in order to create test problems with tractable interaction patterns, we would like to define them such that each problem has only a single lower level optimum for a given $\boldx_u$. To ensure the existence of single lower level optimum, and to enable realistic interactions between the two levels, we consider imposing the following simple restrictions on the objective functions:

\vskip 0.2cm
\textit{Case 1. Creating co-operative interaction: }
A test problem with co-operative interaction pattern can be created by choosing 
\begin{eqnarray}\label{eq:co-op}
F_2(\boldx_{l1}) &=& f_2(\boldx_{l1}), \\
F_3(\boldx_{u2},\boldx_{l2}) &=& F_4(\boldx_{u2})+f_3(\boldx_{u2},\boldx_{l2}) \nonumber,
\end{eqnarray}
where $F_4(\boldx_{u2})$ is any function of $\boldx_{u2}$ whose minimum is known. 
\vskip 0.2cm
\textit{Case 2. Creating conflicting interaction: } 
A test problem with a conflict between the two levels can be created by simply changing the signs of terms $f_2$ and $f_3$  on the right hand side in \eqref{eq:co-op},
\begin{eqnarray}
F_2(\boldx_{l1}) &=& -f_2(\boldx_{l1}), \\
F_3(\boldx_{u2},\boldx_{l2}) &=& F_4(\boldx_{u2})-f_3(\boldx_{u2},\boldx_{l2}). \nonumber
\end{eqnarray}
The choice of $F_2$ and $F_3$ suggested here is a special case, and there can be many other ways to achieve conflict or co-operation using the two functions.
\vskip 0.2cm
\textit{Case 3. Creating mixed interaction: } 
There may be a situation of both cooperation and conflict if functions $F_2$ and $F_3$ are chosen with opposite signs as,
\begin{eqnarray}
F_2(\boldx_{l1}) &=& f_2(\boldx_{l1}),\\
F_3(\boldx_{u2},\boldx_{l2})&=&F_4(\boldx_{u2})-f_3(\boldx_{u2},\boldx_{l2}),\nonumber
\end{eqnarray}
or 
\begin{eqnarray}
F_2(\boldx_{l1})&=& -f_2(\boldx_{l1}),\\
F_3(\boldx_{u2},\boldx_{l2})&=&F_4(\boldx_{u2})+f_3(\boldx_{u2},\boldx_{l2}).\nonumber
\end{eqnarray}
\vskip 0.2cm

\textit{Example 2:} 
Consider a bilevel optimization problem where the lower level task is given by Example 1. According to the above procedures, we can produce a test problem with a conflict between the upper and lower level by defining the upper level objective function as follows,
\begin{equation}
\begin{array}{l}
F_1(\boldx_{u1}) = \sum_{i=1}^{U1} (x_{u1}^{i})^2,\\
F_2(\boldx_{l1}) =  -\sum_{i=1}^{L1} (x_{l1}^{i})^2,\\
F_3(\boldx_{u2}, \boldx_{l2}) = -\sum_{i=1}^{U2} (x_{u2}^{i} - x_{l2}^{i})^2.
\end{array}
\end{equation}
The chosen formulation corresponds to Case 2, where $F_4(\boldx_{u2})=0$. The final optimal solution of the bilevel problem is $F(\boldx_u,\boldx_l)=0$ for $(\boldx_u,\boldx_l)=\boldzero$.

\subsubsection{Multiple Global Solutions at Lower Level}
In this subsection, we discuss constructing test problems with lower level function having multiple global solutions for a given set of upper level variables. To achieve this, we formulate a lower level function which has multiple lower level optima for a given $\boldx_{u}$, such that $\boldx_{l}^{*} \in \Psi(\boldx_{u})$. Then we ensure that out of all these possible lower level optimal solutions one of them ($\boldx_{l}^{**}$) corresponds to the best upper level function value, i.e.,
\begin{equation}
\boldx_{l}^{**} \in \underset{\boldx_{l}^{*}}{\mbox{argmin}} \left\lbrace F(\boldx_{u},\boldx_{l}^{*}) \hspace{2mm} \big| \hspace{2mm} \boldx_{l}^{*} \in \Psi(\boldx_{u}) \right\rbrace.
\end{equation}

To incorporate this difficulty in the problem, we choose the second functions at the upper and lower levels.  Given that the term $f_2(\boldx_{l1})$ is responsible for causing complexities only at the lower level, we can freely formulate it such that it has multiple lower level optimal solutions. From this it necessarily follows that the entire lower level function has multiple optimal solutions.

\vskip 0.2cm
\textit{Example 3:} We describe the construction procedure by considering a simple example, where the cardinalities of the variables are, $dim(\boldx_{u1}) = 2$, $dim(\boldx_{u2}) = 2$, $dim(\boldx_{l1}) = 2$ and $dim(\boldx_{l2}) = 2$, and the lower level function is defined as follows,
\begin{equation}
\begin{array}{l}
f_1(\boldx_{u1}, \boldx_{u2}) = (x_{u1}^{1})^2 + (x_{u1}^{2})^2 + (x_{u2}^{1})^2 + (x_{u2}^{2})^2,\\
f_2(\boldx_{l1}) =   (x_{l1}^{1} - x_{l1}^{2})^2,\\
f_3(\boldx_{u2},\boldx_{l2}) =  (x_{u2}^{1} - x_{l2}^{1})^2 +  (x_{u2}^{2} - x_{l2}^{2})^2.\\
\end{array}
\end{equation}
Here, we observe that $f_2(\boldx_{l1})$ induces multiple optimal solutions, as its minimum value is $0$ for all $x_{l1}^{1} = x_{l1}^{2}$. At the minimum $f_3(\boldx_{u2},\boldx_{l2})$ fixes the values of $x_{l2}^{1}$ and $x_{l2}^{2}$ to $x_{u2}^{1}$ and $x_{u2}^{2}$ respectively. Next, we write the upper level function ensuring that out of the set $x_{l1}^{1} = x_{l1}^{2}$, one of the solutions is best at upper level,
\begin{equation}
\begin{array}{l}
F_1(\boldx_{u1}) = (x_{u1}^{1})^2 + (x_{u1}^{2})^2,\\
F_2(\boldx_{l1}) =   (x_{l1}^{1})^2 + (x_{l1}^{2})^2,\\
F_3(\boldx_{u2},\boldx_{l2}) =  (x_{u2}^{1} - x_{l2}^{2})^2 +  (x_{u2}^{2} - x_{l2}^{2})^2.\\
\end{array}
\end{equation}
The formulation of $F_2(\boldx_{l1})$, as sum of squared terms ensures that $x_{l1}^{1} = x_{l1}^{2} = 0$ provides the best solution at the upper level for any given $(\boldx_{u1},\boldx_{u2})$.

\subsection{Difficulties induced by constraints}
In this subsection, we discuss the types of constraints that can be encountered in a bilevel optimization problem. We only consider inequality constraints in this bilevel test problem construction framework. Considering that the bilevel problems have the possibility to have constraints at both levels, and each constraint could be a function of two different kinds of variables, the constrained set at both levels can be further broken down into smaller subsets as follows:

\begin{table}[h]
\begin{center}
    \begin{tabular}{|c|l|l|l|}
        \hline
Level &        Constraint Set               &       Subsets         & Dependence \\ \hline
Upper      &  $\boldG = \{G_j:j \in J\}$ & $\boldG = \boldG_a \cup \boldG_b \cup \boldG_c$ & $\boldG_a$ depends on $\boldx_u$\\
           &                             &                                                 & $\boldG_b$ depends on $\boldx_l$\\
           &                             &                                                 & $\boldG_c$ depends on $\boldx_u$ and $\boldx_l$\\ \hline
Lower      &  $\boldg = \{g_i:i \in I\}$ & $\boldg = \boldg_a \cup \boldg_b \cup \boldg_c$ & $\boldg_a$ depends on $\boldx_u$\\
           &                             &                                                 & $\boldg_b$ depends on $\boldx_l$\\
           &                             &                                                 & $\boldg_c$ depends on $\boldx_u$ and $\boldx_l$\\ \hline
    \end{tabular}
\caption{Composition of the constraint sets at both levels.}
\label{tab:constraintSets}
\end{center}
\end{table}

In Table~\ref{tab:constraintSets}, $\boldG$ and $\boldg$ denote the set of constraints at the upper and lower level respectively. Each of the constraint set can be broken into three smaller subsets, as shown in the table. The first subset represents constraints that are functions of the upper level variables only, the second subset represents constraints that are functions of lower level variables only, and the third subset represents constraints that are functions of both upper and lower level variables. The reason for splitting the constraints into smaller subsets is to develop an insight for solving these kinds of problems using an evolutionary approach. If the first constraint subset ($\boldG_a$ or $\boldg_a$) is non-empty at either of the two levels, then for any given $\boldx_{u}$ we should check the feasibility of constraints in the sets $\boldG_a$ and $\boldg_a$, before solving the lower level optimization problem. In case, there is one or more infeasible constraints in $\boldg_a$, then the lower level optimization problem does not contain optimal lower level solution ($\boldx_{l}^{*}$) for the given $\boldx_u$. However, if one or more constraints are infeasible within $\boldG_b$, then a lower level optimal solution ($\boldx_{l}^{*}$) may exist for the given $\boldx_u$, but the pair ($\boldx_u,\boldx_{u}^{*}$) will be infeasible for the bilevel problem. Based on this property, a decision can be made, whether it is useful to solve the lower level optimization problem at all for a given $\boldx_u$.

The upper level constraint subsets, $\boldG_b$ depends on $\boldx_l$, and $\boldG_c$ depends on $\boldx_u$ and $\boldx_l$. The values of these constraints are meaningful only when the lower level vector is an optimal solution to the lower level optimization problem. Based on the various constraints which may be functions of $\boldx_u$, or $\boldx_l$ or both, a bilevel problem introduces different kinds of difficulties in the optimization task. In this paper, we aim to construct such examples of constrained bilevel test problems that incorporate some of these complexities. We have proposed four constrained bilevel problems, each of which has at least one or more of the following properties,

\begin{enumerate}
\item Constraints exist but are not active at the optimum
\item A subset of constraints or all the constraints are active at the optimum
\item Upper level constraints are functions of only upper level variables, and lower level constraints are functions of only lower level variables
\item Both upper and lower level constraints are functions of upper as well as lower level variables
\item Lower level constraints lead to multiple global solutions at the lower level
\item Constraints are scalable at both levels
\end{enumerate}

While describing the test problems in the next section, we discuss the construction procedure for the individual constrained test problems.

\section{SMD test problems}\label{sec:testproblems}

By adhering to the design principles introduced in the previous section, we now propose a set of twelve problems which we call as the SMD\footnote{The first six test problems were proposed through a conference publication \citep{my-cec12a}.} test problems. Each problem represents a different difficulty level in terms of convergence at the two levels, complexity of interaction between two levels, and multi-modalities at each of the levels. The first eight problems are unconstrained and the remaining four are constrained.

\subsection{SMD1}

\begin{figure}
\begin{center}
\epsfig{file=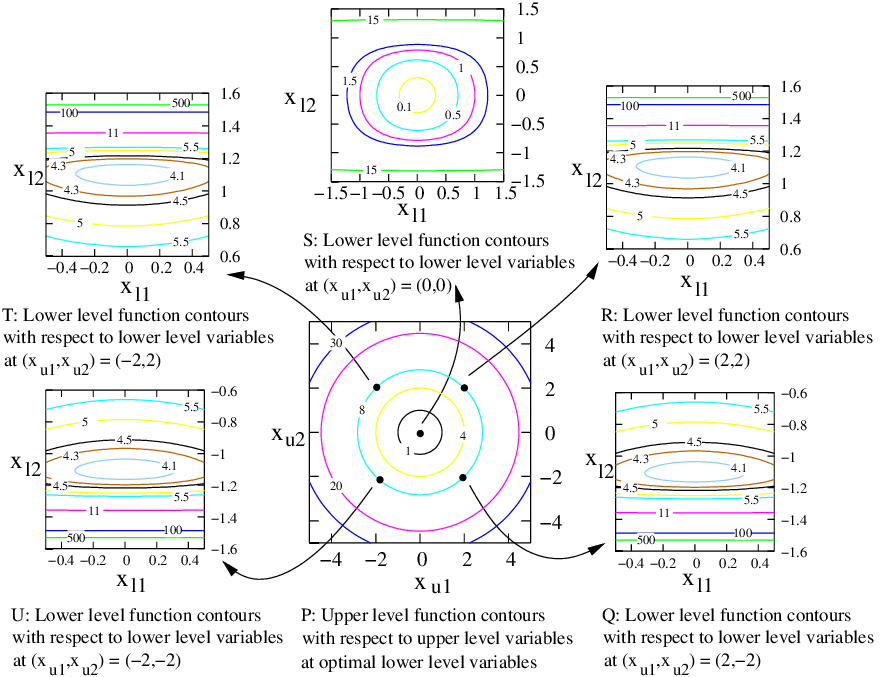,width=0.8\linewidth}
\caption{Upper and lower level function contours for a four-variable SMD1 test problem.}
\label{fig:smd1-2}
\end{center}
\end{figure}

This is a simple test problem, where the lower level problem is a convex optimization task and the upper level is convex with respect to upper level variables and optimal lower level variables. The two levels cooperate with each other. The constituent functions are chosen as
\begin{equation}
\begin{array}{l}
F_1 = \sum_{i=1}^{p} (x_{u1}^{i})^2,\\
F_2 = \sum_{i=1}^{q} (x_{l1}^{i})^2,\\
F_3 = \sum_{i=1}^{r} (x_{u2}^{i})^2 + \sum_{i=1}^{r} (x_{u2}^{i} - \tan x_{l2}^{i})^2,\\
f_1 = \sum_{i=1}^{p} (x_{u1}^{i})^2,\\
f_2 = \sum_{i=1}^{q} (x_{l1}^{i})^2,\\
f_3 = \sum_{i=1}^{r} (x_{u2}^{i} - \tan x_{l2}^{i})^2.\\
\end{array}
\end{equation}
The range of variables is as follows:
\begin{equation}
\begin{array}{l}
x_{u1}^{i} \in [-5,10], \hspace{2mm} \forall \hspace{2mm} i \in \{1,2,\ldots,p\},\\
x_{u2}^{i} \in [-5,10], \hspace{2mm} \forall \hspace{2mm} i \in \{1,2,\ldots,r\},\\
x_{l1}^{i} \in [-5,10], \hspace{2mm} \forall \hspace{2mm} i \in \{1,2,\ldots,q\},\\
x_{l2}^{i} \in (\frac{-\pi}{2},\frac{\pi}{2}), \hspace{2mm} \forall \hspace{2mm} i \in \{1,2,\ldots,r\}.
\end{array}
\end{equation}
Relationship between upper level variables and lower level optimal variables is given as follows:
\begin{equation}
\begin{array}{l}
x_{l1}^{i} = 0, \hspace{2mm} \forall \hspace{2mm} i \in \{1,2,\ldots,p\},\\
x_{l2}^{i} = \tan^{-1} x_{u2}^{i}, \hspace{2mm} \forall \hspace{2mm} i \in \{1,2,\ldots,r\}.
\end{array}
\end{equation}
The values of the variables at the optima are $\boldx_u=0$ and $\boldx_l$ is obtained by the relationship given above. Both upper and lower level functions are equal to zero at the optima.

Figure \ref{fig:smd1-2} shows the contours of the upper and lower level functions with respect to the upper and lower level variables for a four-variable test problem. The problem has two upper level variables and two lower level variables, such that the dimensions of $\boldx_{u1}, \boldx_{u2}, \boldx_{l1}$ and $\boldx_{u2}$ are all one. Sub-figure P shows the upper level function contours with respect to the upper level variables, assuming that the lower level variables are at the optima. Fixing the upper level variables $(\boldx_{u1},\boldx_{u2})$ at five different locations, i.e. $(2,2), (-2,2), (2,-2), (-2,-2)$ and $(0,0)$, the lower level function contours are shown with respect to the lower level variables. This shows that the contours of the lower level optimization problem may be different for different upper level vectors.

\begin{figure}
\begin{center}
\epsfig{file=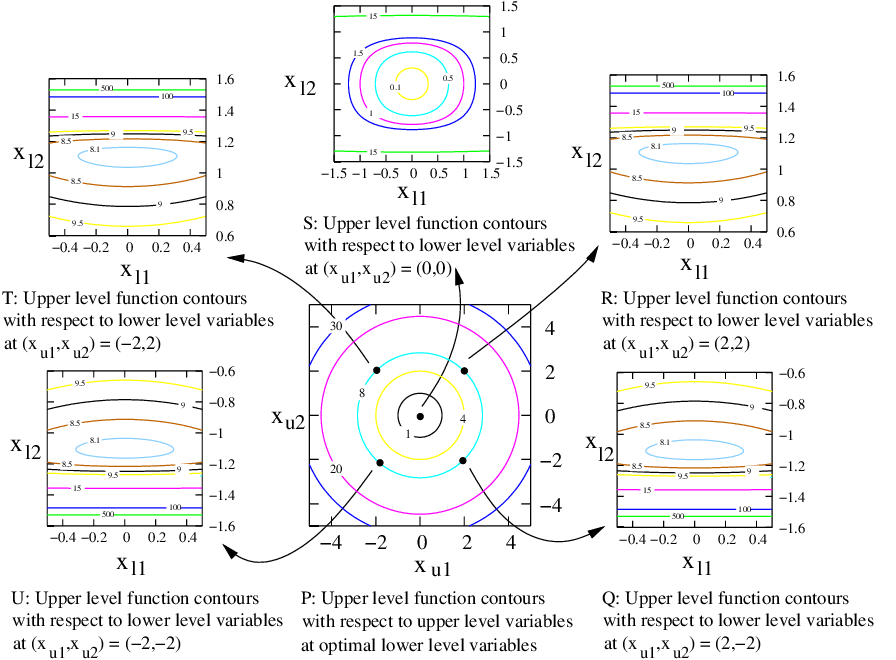,width=0.8\linewidth}
\caption{Upper level function contours for a four-variable SMD1 test problem.}
\label{fig:smd1}
\end{center}
\end{figure}

Figure \ref{fig:smd1} shows the contours of the upper level function with respect to the upper and lower level variables. Sub-figure P once again shows the upper level function contours with respect to the upper level variables. However, sub-figures Q, R, S, T and V now represent the upper level function contours at different $(\boldx_{u1},\boldx_{u2})$, i.e. $(2,2), (-2,2), (2,-2), (-2,-2)$ and $(0,0)$. From sub-figures Q, R, S, T and V, we observe that if the lower level variables move away from its optimal location, the upper level function value deteriorates. This means that the upper level function and the lower level functions are cooperative.

\subsection{SMD2}

\begin{figure}
\begin{center}
\epsfig{file=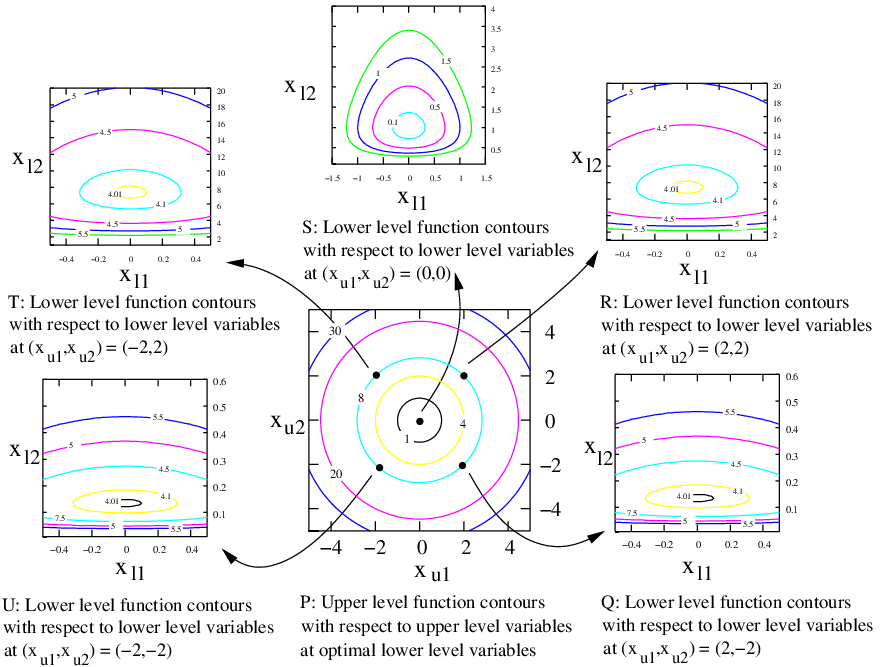,width=0.8\linewidth}
\caption{Upper and lower level function contours for a four-variable SMD2 test problem.}
\label{fig:smd2-2}
\end{center}
\end{figure}

This test problem is similar to the SMD1 test problem. However, there is a conflict between the upper level and lower level optimization task. The lower level optimization problem is once again a convex optimization task and the upper level optimization is convex with respect to upper level variables and optimal lower level variables. Since the two levels are conflicting, an inaccurate lower level optimum may lead to upper level function value better than the true optimum for the bilevel problem. The constituent functions are chosen as
\begin{equation}
\begin{array}{l}
F_1 = \sum_{i=1}^{p} (x_{u1}^{i})^2,\\
F_2 = - \sum_{i=1}^{q} (x_{l1}^{i})^2,\\
F_3 = \sum_{i=1}^{r} (x_{u2}^{i})^2 - \sum_{i=1}^{r} (x_{u2}^{i} - \log x_{l2}^{i})^2,\\
f_1 = \sum_{i=1}^{p} (x_{u1}^{i})^2,\\
f_2 = \sum_{i=1}^{q} (x_{l1}^{i})^2,\\
f_3 = \sum_{i=1}^{r} (x_{u2}^{i} - \log x_{l2}^{i})^2.\\
\end{array}
\end{equation}
The range of variables is as follows:
\begin{equation}
\begin{array}{l}
x_{u1}^{i} \in [-5,10], \hspace{2mm} \forall \hspace{2mm} i \in \{1,2,\ldots,p\},\\
x_{u2}^{i} \in [-5,1], \hspace{2mm} \forall \hspace{2mm} i \in \{1,2,\ldots,r\},\\
x_{l1}^{i} \in [-5,10], \hspace{2mm} \forall \hspace{2mm} i \in \{1,2,\ldots,q\},\\
x_{l2}^{i} \in (0,e], \hspace{2mm} \forall \hspace{2mm} i \in \{1,2,\ldots,r\}.
\end{array}
\end{equation}
Relationship between upper level variables and lower level optimal variables is given as follows:
\begin{equation}
\begin{array}{l}
x_{l1}^{i} = 0, \hspace{2mm} \forall \hspace{2mm} i \in \{1,2,\ldots,q\},\\
x_{l2}^{i} = \log^{-1} x_{u2}^{i}, \hspace{2mm} \forall \hspace{2mm} i \in \{1,2,\ldots,r\}.
\end{array}
\end{equation}
The values of the variables at the optima are $\boldx_u=0$ and $\boldx_l$ is obtained by the relationship given above. Both upper and lower level functions are equal to zero at the optima.

Figure \ref{fig:smd2-2} shows the contours of the upper and lower level functions with respect to the upper and lower level variables for a four-variable test problem. The problem has two upper level variables and two lower level variables, such that the dimension of $\boldx_{u1}, \boldx_{u2}, \boldx_{l1}$ and $\boldx_{u2}$ are all one. The figure provides the same information about SMD2, as Figure \ref{fig:smd1-2} provides about SMD1. However, the shape of the contours differ, which is caused by the use of different $F_3$ and $f_3$ functions.

\begin{figure}
\begin{center}
\epsfig{file=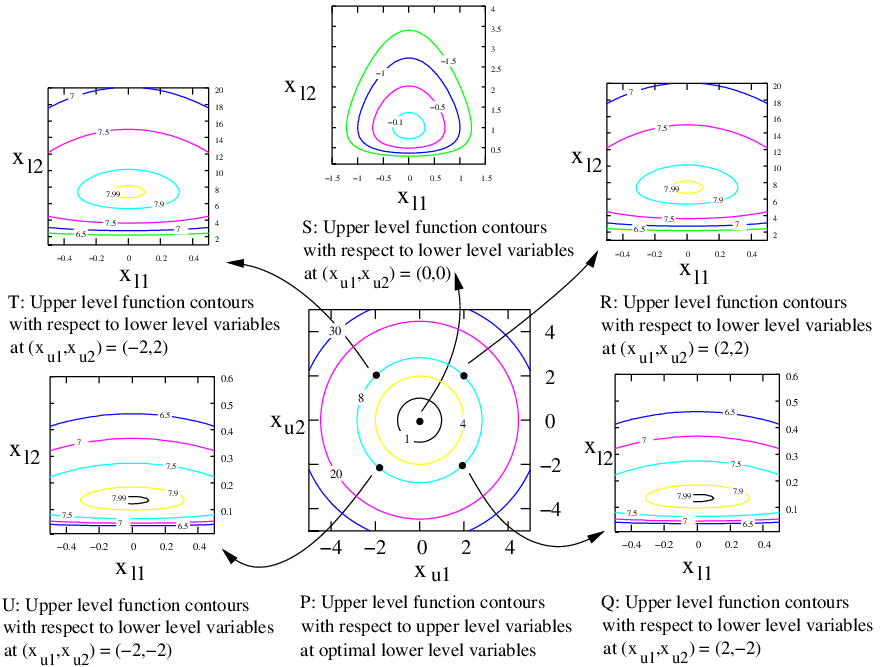,width=0.8\linewidth}
\caption{Upper level function contours for a four-variable SMD2 test problem.}
\label{fig:smd2}
\end{center}
\end{figure}

Figure \ref{fig:smd2} shows the contours of the upper level function with respect to the upper and lower level variables, and provides the same information as Figure \ref{fig:smd1} provides about SMD1. This figure shows the conflicting nature of the problem caused by using a negative sign in $F_2$. The conflicting nature can be observed from the sub-figures Q, R, S, T and U. For a given $\boldx_u$, as one moves away from the lower level optimal solution, the upper level function value further reduces. On the other hand, in Figure \ref{fig:smd2-2} we observe that moving away from the lower level optimal solution causes an increase in lower level function value.

\subsection{SMD3}
In this test problem there is a cooperation between the two levels. The difficulty is introduced in terms of multi-modality at the lower level which contains the Rastrigin's function. The upper level is convex with respect to upper level variables and optimal lower level variables. The constituent functions are chosen as
\begin{equation}
\begin{array}{l}
F_1 = \sum_{i=1}^{p} (x_{u1}^{i})^2,\\
F_2 = \sum_{i=1}^{q} (x_{l1}^{i})^2,\\
F_3 = \sum_{i=1}^{r} (x_{u2}^{i})^2 + \sum_{i=1}^{r} ((x_{u2}^{i})^2 - \tan x_{l2}^{i})^2,\\
f_1 = \sum_{i=1}^{p} (x_{u1}^{i})^2,\\
f_2 = q + \sum_{i=1}^{q} \left(\left(x_{l1}^{i}\right)^2 - \cos 2\pi x_{l1}^{i}\right),\\
f_3 = \sum_{i=1}^{r} ((x_{u2}^{i})^2 - \tan x_{l2}^{i})^2.\\
\end{array}
\end{equation}
The range of variables is as follows:
\begin{equation}
\begin{array}{l}
x_{u1}^{i} \in [-5,10], \hspace{2mm} \forall \hspace{2mm} i \in \{1,2,\ldots,p\},\\
x_{u2}^{i} \in [-5,10], \hspace{2mm} \forall \hspace{2mm} i \in \{1,2,\ldots,r\},\\
x_{l1}^{i} \in [-5,10], \hspace{2mm} \forall \hspace{2mm} i \in \{1,2,\ldots,q\},\\
x_{l2}^{i} \in (\frac{-\pi}{2},\frac{\pi}{2}), \hspace{2mm} \forall \hspace{2mm} i \in \{1,2,\ldots,r\}.
\end{array}
\end{equation}
Relationship between upper level variables and lower level optimal variables is given as follows:
\begin{equation}
\begin{array}{l}
x_{l1}^{i} = 0, \hspace{2mm} \forall \hspace{2mm} i \in \{1,2,\ldots,q\},\\
x_{l2}^{i} = \tan^{-1} (x_{u2}^{i})^2, \hspace{2mm} \forall \hspace{2mm} i \in \{1,2,\ldots,r\}.
\end{array}
\end{equation}
The values of the variables at the optima are $\boldx_u=0$ and $\boldx_l$ is obtained by the relationship given above. Both upper and lower level functions are equal to zero at the optima. Rastrigin's function used in $f_2$ has multiple local optima around the global optimum, which introduces convergence difficulties at the lower level.

\begin{figure}
\begin{center}
\epsfig{file=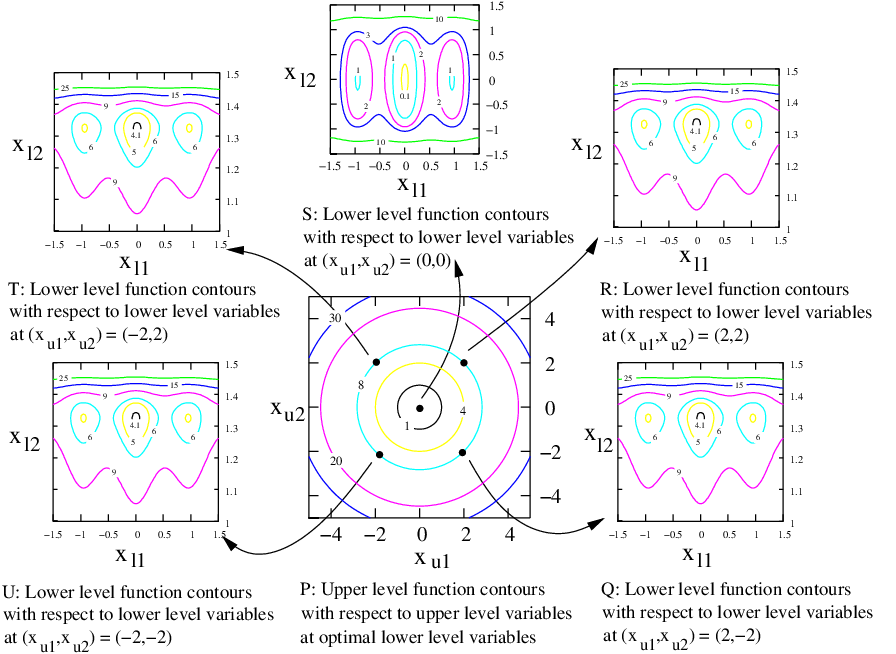,width=0.8\linewidth}
\caption{Upper and lower level function contours for a four-variable SMD3 test problem.}
\label{fig:smd3}
\end{center}
\end{figure}

Sub-figure P in Figure \ref{fig:smd3} shows the contours of the upper level function with respect to the upper level variables assuming the lower level variables to be optimal at each $\boldx_u$. Sub-figures Q, R, S, T, and U show the behavior of the lower level function at 5 different locations of $\boldx_u$, which are $(2,2), (-2,2), (2,-2), (-2,-2)$ and $(0,0)$. The problem is once again assumed to have two upper level variables and two lower level variables, such that the dimensions of $\boldx_{u1}, \boldx_{u2}, \boldx_{l1}$ and $\boldx_{u2}$ are all one. The figure shows that there is a different lower level optimization problem at each $\boldx_u$ which is required to be solved in order to achieve a feasible solution at the upper level. The contours of the lower level optimization problem differ based on the location of upper level vector. It can be observed that the Rastrigin's function at the lower level introduces multiple local optima into the problem. The contours of the lower level are further distorted because of the presence of the $\tan (\cdot)$ function at the lower level.

In spite of multiple local optima at the lower level, this problem is easier to solve because of the cooperating nature of the functions at the two levels. If a lower level optimization problem is stuck at a local optimum for a particular $\boldx_u$ (say $\boldx_{u}^{(0)}$), it will have a poorer objective function value at the upper level. However, as soon as another lower level optimization problem is solved in the vicinity of $\boldx_{u}^{(0)}$, which attains a global lower level optimum, then it will have a better objective function value at the upper level and will dominate the previous inaccurate solution. Therefore, a method which is able to handle multi-modality at the lower level at least in few of its lower level optimization runs will be able to successfully solve this problem.

\subsection{SMD4}
In this test problem there is a conflict between the two levels. The difficulty is in terms of multi-modality at the lower level which once again contains the Rastrigin's function. The upper level is convex with respect to upper level variables and optimal lower level variables. The constituent functions are chosen as
\begin{equation}
\begin{array}{l}
F_1 = \sum_{i=1}^{p} (x_{u1}^{i})^2,\\
F_2 = - \sum_{i=1}^{q} (x_{l1}^{i})^2,\\
F_3 = \sum_{i=1}^{r} (x_{u2}^{i})^2 - \sum_{i=1}^{r} (|x_{u2}^{i}| - \log (1+x_{l2}^{i}))^2,\\
f_1 = \sum_{i=1}^{p} (x_{u1}^{i})^2,\\
f_2 = q + \sum_{i=1}^{q} \left(\left(x_{l1}^{i}\right)^2 - \cos 2\pi x_{l1}^{i}\right),\\
f_3 = \sum_{i=1}^{r} (|x_{u2}^{i}| - \log(1+x_{l2}^{i}))^2.\\
\end{array}
\end{equation}
The range of variables is as follows:
\begin{equation}
\begin{array}{l}
x_{u1}^{i} \in [-5,10], \hspace{2mm} \forall \hspace{2mm} i \in \{1,2,\ldots,p\},\\
x_{u2}^{i} \in [-1,1], \hspace{2mm} \forall \hspace{2mm} i \in \{1,2,\ldots,r\},\\
x_{l1}^{i} \in [-5,10], \hspace{2mm} \forall \hspace{2mm} i \in \{1,2,\ldots,q\},\\
x_{l2}^{i} \in [0,e], \hspace{2mm} \forall \hspace{2mm} i \in \{1,2,\ldots,r\}.
\end{array}
\end{equation}
Relationship between upper level variables and lower level optimal variables is given as follows:
\begin{equation}
\begin{array}{l}
x_{l1}^{i} = 0, \hspace{2mm} \forall \hspace{2mm} i \in \{1,2,\ldots,q\},\\
x_{l2}^{i} = \log^{-1} |x_{u2}^{i}| - 1, \hspace{2mm} \forall \hspace{2mm} i \in \{1,2,\ldots,r\}.
\end{array}
\end{equation}
The values of the variables at the optima are $\boldx_u=0$ and $\boldx_l$ is obtained by the relationship given above. Both upper and
lower level functions are equal to zero at the optima.

\begin{figure}
\begin{center}
\epsfig{file=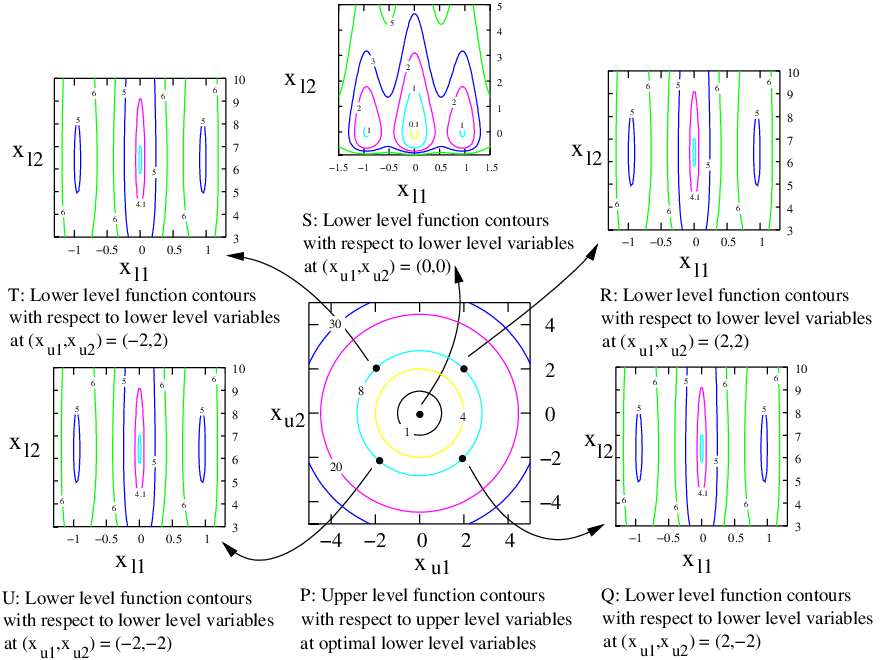,width=0.8\linewidth}
\caption{Upper and lower level function contours for a four-variable SMD4 test problem.}
\label{fig:smd4}
\end{center}
\end{figure}

Figure~\ref{fig:smd4} represents the same information as in Figure~\ref{fig:smd3} for a four-variable bilevel problem. However, this problem involves conflict between the two levels, which makes it significantly more difficult to solve than the previous test problem. If a lower level optimization problem is stuck at a local optimum for a particular $\boldx_u$, it will end up having a better objective function value at the upper level than what it will attain at the true global lower level optimum. Therefore, even if another lower level optimization problem is successfully solved in the vicinity of $\boldx_{u}$, the previous inaccurate solution will dominate the new solution at the upper level. This problem can be handled only by those methods which are able to efficiently handle lower level multi-modality without getting stuck in a local basin.

\subsection{SMD5}
In this test problem, there is a conflict between the two levels. The difficulty introduced is in terms of multi-modality and convergence at the lower level. The lower level problem contains the Rosenbrock's (banana) function such that the global optimum lies in a long, narrow, flat parabolic valley. 
The upper level is convex with respect to upper level variables and optimal lower level variables. The constituent functions are chosen as
\begin{equation}
\begin{array}{l}
F_1 = \sum_{i=1}^{p} (x_{u1}^{i})^2,\\
F_2 = - \sum_{i=1}^{q-1} \left( \left(x_{l1}^{i+1} - \left(x_{l1}^{i}\right)^2\right)^2 + \left(x_{l1}^{i} - 1\right)^2 \right),\\
F_3 = \sum_{i=1}^{r} (x_{u2}^{i})^2 - \sum_{i=1}^{r} (|x_{u2}^{i}| - (x_{l2}^{i})^2)^2,\\
f_1 = \sum_{i=1}^{p} (x_{u1}^{i})^2,\\
f_2 = \sum_{i=1}^{q-1} \left( \left(x_{l1}^{i+1} - \left(x_{l1}^{i}\right)^2\right)^2 + \left(x_{l1}^{i} - 1\right)^2 \right),\\
f_3 = \sum_{i=1}^{r} (|x_{u2}^{i}| - (x_{l2}^{i})^2)^2.
\end{array}
\end{equation}
The range of variables is as follows:
\begin{equation}
\begin{array}{l}
x_{u1}^{i} \in [-5,10], \hspace{2mm} \forall \hspace{2mm} i \in \{1,2,\ldots,p\},\\
x_{u2}^{i} \in [-5,10], \hspace{2mm} \forall \hspace{2mm} i \in \{1,2,\ldots,r\},\\
x_{l1}^{i} \in [-5,10], \hspace{2mm} \forall \hspace{2mm} i \in \{1,2,\ldots,q\},\\
x_{l2}^{i} \in [-5,10], \hspace{2mm} \forall \hspace{2mm} i \in \{1,2,\ldots,r\}.
\end{array}
\end{equation}
Relationship between upper level variables and lower level optimal variables is given as follows:
\begin{equation}
\begin{array}{l}
x_{l1}^{i} = 1, \hspace{2mm} \forall \hspace{2mm} i \in \{1,2,\ldots,q\},\\
x_{l2}^{i} = \sqrt{|x_{u2}^{i}|}, \hspace{2mm} \forall \hspace{2mm} i \in \{1,2,\ldots,r\}.
\end{array}
\end{equation}
The values of the variables at the optima are $\boldx_u=0$ and $\boldx_l$ is obtained by the relationship given above. Both upper and
lower level functions are equal to zero at the optima.

\subsection{SMD6}
In this test problem, there is again a conflict between the two levels. However, this problem differs from the previous problems by containing infinitely many global solutions at the lower level for any given upper level vector. Out of the entire global solution set, there is only a single lower level point which corresponds to the best upper level function value. The constituent functions are chosen as
\begin{equation}
\begin{array}{l}
F_1 = \sum_{i=1}^{p} (x_{u1}^{i})^2,\\
F_2 = - \sum_{i=1}^{q} (x_{l1}^{i})^2 + \sum_{i=q+1}^{q+s} (x_{l1}^{i})^2,\\
F_3 = \sum_{i=1}^{r} (x_{u2}^{i})^2 - \sum_{i=1}^{r} (x_{u2}^{i} - x_{l2}^{i})^2,\\
f_1 = \sum_{i=1}^{p} (x_{u1}^{i})^2,\\
f_2 = \sum_{i=1}^{q} (x_{l1}^{i})^2 + \sum_{i=q+1, i=i+2}^{q+s-1} (x_{l1}^{i+1} - x_{l1}^{i})^2,\\
f_3 = \sum_{i=1}^{r} (x_{u2}^{i} - x_{l2}^{i})^2.
\end{array}
\end{equation}
The range of variables is as follows:
\begin{equation}
\begin{array}{l}
x_{u1}^{i} \in [-5,10], \hspace{2mm} \forall \hspace{2mm} i \in \{1,2,\ldots,p\},\\
x_{u2}^{i} \in [-5,10], \hspace{2mm} \forall \hspace{2mm} i \in \{1,2,\ldots,r\},\\
x_{l1}^{i} \in [-5,10], \hspace{2mm} \forall \hspace{2mm} i \in \{1,2,\ldots,q+s\},\\
x_{l2}^{i} \in [-5,10], \hspace{2mm} \forall \hspace{2mm} i \in \{1,2,\ldots,r\}.
\end{array}
\end{equation}
Relationship between upper level variables and lower level optimal variables is given as follows:
\begin{equation}
\begin{array}{l}
x_{l1}^{i} = 0, \hspace{2mm} \forall \hspace{2mm} i \in \{1,2,\ldots,q\},\\
x_{l2}^{i} = x_{u2}^{i}, \hspace{2mm} \forall \hspace{2mm} i \in \{1,2,\ldots,r\}.
\end{array}
\end{equation}
The values of the variables at the optima are $\boldx_u=0$ and $\boldx_l$ is obtained by the relationship given above. Both upper and
lower level functions are equal to zero at the optima.

\begin{figure}
\begin{center}
\epsfig{file=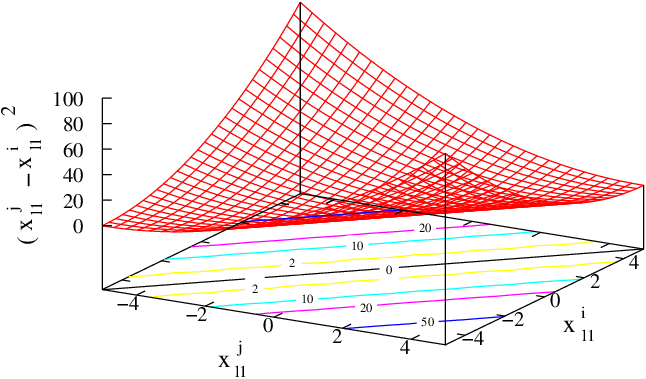,width=0.7\linewidth}
\caption{Plot of the term in $f_2$ responsible for creating multiple optimum solutions at the lower level. The value of the term is zero at all the points in the valley.}
\label{fig:multipleOpt}
\end{center}
\end{figure}

Figure~\ref{fig:multipleOpt} shows the second term ($(x_{l1}^{i} - x_{l1}^{j})^2$, for $s=2$) for function $f_2$, and its contours at the lower level. It can be observed from the figure that all the points along $\boldx_{l1}^{j} = \boldx_{l2}^{i}$ have a value 0 for the function $f_2$. All these points are responsible for introducing multiple global optimal solutions at the lower level for any given upper level variable vector. However, out of all the global optimal solutions at the lower level, the solution $\boldx_{l1}^{j} = \boldx_{l2}^{i} = 0$ provides the best function value at the upper level for any given upper level variable vector.

\subsection{SMD7}
In this test problem, we introduce complexities at the upper level while keeping the lower level optimization task relatively simpler. Most of the previous test problems would be useful for testing the ability of algorithms to handle lower level optimization task efficiently. However, this test problem contains multi-modality at the upper level, which demands a global optimization approach at the upper level. The function $F_1$ at the upper level represents a slightly modified Griewank function. The constituent functions are chosen as

\begin{equation}
\begin{array}{l}
F_1 = 1 + \frac{1}{400}\sum_{i=1}^{p} \left(x_{u1}^{i}\right)^2 - \Pi_{i=1}^{p} \left(\cos \frac{x_{u1}^{i}}{\sqrt{i}}\right),\\
F_2 = - \sum_{i=1}^{q} (x_{l1}^{i})^2,\\
F_3 = \sum_{i=1}^{r} (x_{u2}^{i})^2 - \sum_{i=1}^{r} (x_{u2}^{i} - \log x_{l2}^{i})^2,\\
f_1 = \sum_{i=1}^{p} (x_{u1}^{i})^3,\\
f_2 = \sum_{i=1}^{q} (x_{l1}^{i})^2,\\
f_3 = \sum_{i=1}^{r} (x_{u2}^{i} - \log x_{l2}^{i})^2.\\
\end{array}
\end{equation}
The range of variables is as follows:
\begin{equation}
\begin{array}{l}
x_{u1}^{i} \in [-5,10], \hspace{2mm} \forall \hspace{2mm} i \in \{1,2,\ldots,p\},\\
x_{u2}^{i} \in [-5,1], \hspace{2mm} \forall \hspace{2mm} i \in \{1,2,\ldots,r\},\\
x_{l1}^{i} \in [-5,10], \hspace{2mm} \forall \hspace{2mm} i \in \{1,2,\ldots,q\},\\
x_{l2}^{i} \in (0,e], \hspace{2mm} \forall \hspace{2mm} i \in \{1,2,\ldots,r\}.
\end{array}
\end{equation}
Relationship between upper level variables and lower level optimal variables is given as follows:
\begin{equation}
\begin{array}{l}
x_{l1}^{i} = 0, \hspace{2mm} \forall \hspace{2mm} i \in \{1,2,\ldots,q\},\\
x_{l2}^{i} = \log^{-1} x_{u2}^{i}, \hspace{2mm} \forall \hspace{2mm} i \in \{1,2,\ldots,r\}.
\end{array}
\end{equation}
The values of the variables at the optima are $\boldx_u=0$ and $\boldx_l$ is obtained by the relationship given above. Both upper and
lower level functions are equal to zero at the optima.

\subsection{SMD8}
This test problem tests the ability of the algorithms to handle multi-modality at the upper level, and convergence complexity at lower level at the same time. There is also a conflict between the upper level and lower level optimization tasks. The lower level objective contains the Rosenbrock's (banana) function, and the upper level objective contains the multi-modal Ackley's function. The constituent functions are chosen as
\begin{equation}
\begin{array}{l}
F_1 = 20 + e -20 exp\left( -0.2 \sqrt{\frac{1}{p}\sum_{i=1}^{p} (x_{u1}^{i})^2}\right) - exp\left(\frac{1}{p}\sum_{i=1}^{p} \cos 2 \pi x_{u1}^{i}\right),\\
F_2 = - \sum_{i=1}^{q-1} \left( \left(x_{l1}^{i+1} - \left(x_{l1}^{i}\right)^2\right)^2 + \left(x_{l1}^{i} - 1\right)^2 \right),\\
F_3 = \sum_{i=1}^{r} (x_{u2}^{i})^2 - \sum_{i=1}^{r} (x_{u2}^{i} - (x_{l2}^{i})^3)^2,\\
f_1 = \sum_{i=1}^{p} |x_{u1}^{i}|,\\
f_2 = \sum_{i=1}^{q-1} \left( \left(x_{l1}^{i+1} - \left(x_{l1}^{i}\right)^2\right)^2 + \left(x_{l1}^{i} - 1\right)^2 \right),\\
f_3 = \sum_{i=1}^{r} (x_{u2}^{i} - (x_{l2}^{i})^3)^2.
\end{array}
\end{equation}
The range of variables is as follows:
\begin{equation}
\begin{array}{l}
x_{u1}^{i} \in [-5,10], \hspace{2mm} \forall \hspace{2mm} i \in \{1,2,\ldots,p\},\\
x_{u2}^{i} \in [-5,10], \hspace{2mm} \forall \hspace{2mm} i \in \{1,2,\ldots,r\},\\
x_{l1}^{i} \in [-5,10], \hspace{2mm} \forall \hspace{2mm} i \in \{1,2,\ldots,q\},\\
x_{l2}^{i} \in [-5,10], \hspace{2mm} \forall \hspace{2mm} i \in \{1,2,\ldots,r\}.
\end{array}
\end{equation}
Relationship between upper level variables and lower level optimal variables is given as follows:
\begin{equation}
\begin{array}{l}
x_{l1}^{i} = 1, \hspace{2mm} \forall \hspace{2mm} i \in \{1,2,\ldots,q\},\\
x_{l2}^{i} = (x_{u2}^{i})^{\frac{1}{3}}, \hspace{2mm} \forall \hspace{2mm} i \in \{1,2,\ldots,r\}.
\end{array}
\end{equation}
The values of the variables at the optima are $\boldx_u=0$ and $\boldx_l$ is obtained by the relationship given above. Both upper and lower level functions are equal to zero at the optima.

\subsection{SMD9}
\begin{figure}
\begin{center}
\epsfig{file=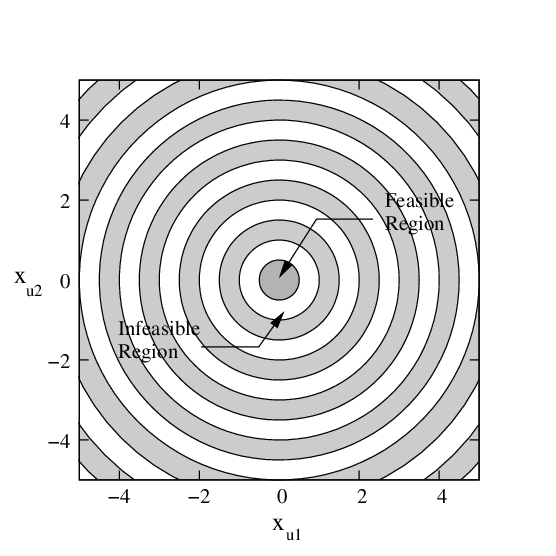,width=0.5\linewidth}
\caption{Feasible and infeasible regions in case of a two-variable constraint function.}
\label{fig:annular}
\end{center}
\end{figure}

In this test problem, we introduce constraints at both upper and lower levels. Constraints are defined such that they cause convergence difficulties at both levels independently. One constraint is introduced at each level, such that the upper level constraint is a function of the upper level variables and the lower level constraint is a function of the lower level variables. The constraints divide the search space into annular regions, and cause convergence difficulties without altering the global optimum. The constraint at the upper as well as the lower level are however, inactive at the optimum. The two levels are once again conflicting in nature, such that an inaccurate lower level optimum may lead to upper level function value better than the true optimum for the bilevel problem. The constituent functions are chosen as
\begin{equation}
\begin{array}{l}
F_1 = \sum_{i=1}^{p} (x_{u1}^{i})^2,\\
F_2 = - \sum_{i=1}^{q} (x_{l1}^{i})^2,\\
F_3 = \sum_{i=1}^{r} (x_{u2}^{i})^2 - \sum_{i=1}^{r} (x_{u2}^{i} - \log (1+x_{l2}^{i}))^2,\\
f_1 = \sum_{i=1}^{p} (x_{u1}^{i})^2,\\
f_2 = \sum_{i=1}^{q} (x_{l1}^{i})^2,\\
f_3 = \sum_{i=1}^{r} (x_{u2}^{i} - \log (1+x_{l2}^{i}))^2.\\
\end{array}
\end{equation}
The upper and lower level constraints are as follows:
\begin{equation}
\begin{array}{l}
\mbox{Upper level constraint}\\
G_1: \frac{\sum_{i=1}^{p} (x_{u1}^{i})^2 + \sum_{i=1}^{r} (x_{u2}^{i})^2}{a} - \Big\lfloor \frac{\sum_{i=1}^{p} (x_{u1}^{i})^2 + \sum_{i=1}^{r} (x_{u2}^{i})^2}{a} + \frac{0.5}{b}\Big\rfloor \ge0,\\
\mbox{Lower level constraint}\\
g_1: \frac{\sum_{i=1}^{p} (x_{l1}^{i})^2 + \sum_{i=1}^{r} (x_{l2}^{i})^2}{a} - \Big\lfloor \frac{\sum_{i=1}^{p} (x_{l1}^{i})^2 + \sum_{i=1}^{r} (x_{l2}^{i})^2}{a} + \frac{0.5}{b}\Big\rfloor \ge0,\\
\mbox{where } a = 1 \mbox{ and } b = 1.
\end{array}
\end{equation}
The range of variables is as follows:
\begin{equation}
\begin{array}{l}
x_{u1}^{i} \in [-5,10], \hspace{2mm} \forall \hspace{2mm} i \in \{1,2,\ldots,p\},\\
x_{u2}^{i} \in [-5,1], \hspace{2mm} \forall \hspace{2mm} i \in \{1,2,\ldots,r\},\\
x_{l1}^{i} \in [-5,10], \hspace{2mm} \forall \hspace{2mm} i \in \{1,2,\ldots,q\},\\
x_{l2}^{i} \in (-1,-1+e], \hspace{2mm} \forall \hspace{2mm} i \in \{1,2,\ldots,r\}.
\end{array}
\end{equation}
Relationship between upper level variables (feasible with respect to upper level constraints) and lower level optimal variables is given as follows:
\begin{equation}
\begin{array}{l}
x_{l1}^{i} = 0, \hspace{2mm} \forall \hspace{2mm} i \in \{1,2,\ldots,q\},\\
x_{l2}^{i} = \log^{-1} x_{u2}^{i} - 1, \hspace{2mm} \forall \hspace{2mm} i \in \{1,2,\ldots,r\}.
\end{array}
\end{equation}
Figure \ref{fig:annular} shows the restricted search space for the upper level optimization task when it is a function of two upper level variables, i.e. $p=1$ and $r=1$. The search space looks similar at the lower level when $q=1$ and $r=1$. For higher number of variables, the annular rings transform into spherical shells. The values of the variables at the optima are $\boldx_u=0$ and $\boldx_l=0$. Both upper and
lower level functions are equal to zero at the optima.

\subsection{SMD10}
In this test problem, we introduce constraints at the upper as well as the lower level such that they are scalable. As the number of variables are varied at the upper and the lower levels, the number of constraints also vary. This is different from the previous problem such that all the constraints are active at the optimum. However, in this case we have the upper level constraints as functions of the upper level variables, and the lower level constraints as functions of the lower level variables. The constituent functions are chosen as
\begin{equation}
\begin{array}{l}
F_1 = \sum_{i=1}^{p} (x_{u1}^{i} - 2)^2,\\
F_2 = \sum_{i=1}^{q} (x_{l1}^{i})^2,\\
F_3 = \sum_{i=1}^{r} (x_{u2}^{i} - 2)^2 - \sum_{i=1}^{r} (x_{u2}^{i} - \tan x_{l2}^{i})^2,\\
f_1 = \sum_{i=1}^{p} (x_{u1}^{i})^2,\\
f_2 = \sum_{i=1}^{q} (x_{l1}^{i} - 2)^2,\\
f_3 = \sum_{i=1}^{r} (x_{u2}^{i} - \tan x_{l2}^{i})^2.\\
\end{array}
\end{equation}
The upper and lower level constraints are as follows:
\begin{equation}
\begin{array}{l}
\mbox{Upper level constraints}\\
G_j: x_{u1}^{j} - \sum_{i=1,i \ne j}^{p} (x_{u1}^{i})^3 - \sum_{i=1}^{r} (x_{u2}^{i})^3  \ge 0, \hspace{1mm} \forall \hspace{1mm} j \in \{1,2,\ldots,p\},\\
G_{p+j}: x_{u2}^{j} - \sum_{i=1,i \ne j}^{r} (x_{u2}^{i})^3 - \sum_{i=1}^{p} (x_{u1}^{i})^3  \ge 0, \hspace{1mm} \forall \hspace{1mm} j \in \{1,2,\ldots,r\},\\
\mbox{Lower level constraints}\\
g_j: x_{l1}^{j} - \sum_{i=1,i \ne j}^{q} (x_{l1}^{i})^3  \ge 0, \hspace{1mm} \forall \hspace{1mm} j \in \{1,2,\ldots,q\}.\\
\end{array}
\end{equation}
The range of variables is as follows:
\begin{equation}
\begin{array}{l}
x_{u1}^{i} \in [-5,10], \hspace{2mm} \forall \hspace{2mm} i \in \{1,2,\ldots,p\},\\
x_{u2}^{i} \in [-5,10], \hspace{2mm} \forall \hspace{2mm} i \in \{1,2,\ldots,r\},\\
x_{l1}^{i} \in [-5,10], \hspace{2mm} \forall \hspace{2mm} i \in \{1,2,\ldots,q\},\\
x_{l2}^{i} \in (\frac{-\pi}{2},\frac{\pi}{2}), \hspace{2mm} \forall \hspace{2mm} i \in \{1,2,\ldots,r\}.
\end{array}
\end{equation}
Relationship between upper level variables (feasible with respect to upper level constraints) and lower level optimal variables is given as follows:
\begin{equation}
\begin{array}{l}
x_{l1}^{i} = \frac{1}{\sqrt{q-1}}, \hspace{2mm} \forall \hspace{2mm} i \in \{1,2,\ldots,q\},\\
x_{l2}^{i} = \tan^{-1} x_{u2}^{i}, \hspace{2mm} \forall \hspace{2mm} i \in \{1,2,\ldots,r\}.
\end{array}
\end{equation}
The values of the variables at the optima are $\boldx_u=\frac{1}{\sqrt{p+r-1}}$, and $\boldx_l$ is obtained by the relationship given above.

\begin{figure}
\begin{center}
\epsfig{file=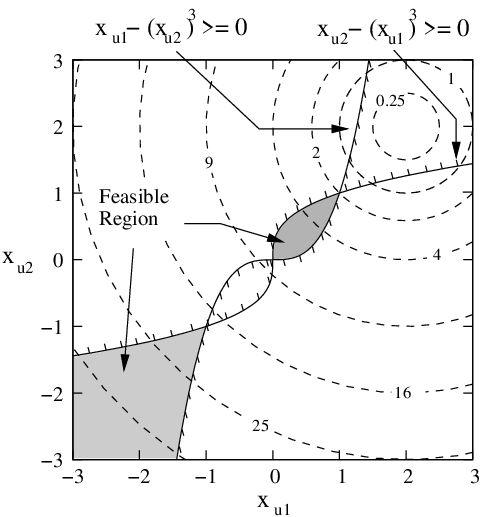,width=0.45\linewidth}
\caption{Feasible and infeasible regions in case of a two-variable constraint function.}
\label{fig:cubicConst}
\end{center}
\end{figure}

Figure \ref{fig:cubicConst} shows the feasible region of the search space for the upper level optimization task, when the upper level objective is a function of two upper variables, i.e. $p=1, r=1$. The shaded part in the figure shows the feasible region, and the dotted lines show the contours of the upper level objective function. For the given two variable upper level objective function, the optima lies at one of the intersections ($(\boldx_{u1}, \boldx_{u2})=(1,1)$) of the constraints shown in the figure.

\subsection{SMD11}
In this test problem, we introduce constraints that are functions of upper as well as lower variables at both levels. The constraints at the upper level are scalable, but there is just a single constraint at the lower level. The constraint at the lower level introduces multiple global optimal solutions at the lower level for any given upper level vector. At the optimum of the bilevel problem, the lower level constraint as well as the upper level constraints are active. The upper level constraints eliminate a large part of the global optimal solutions from the lower level. The constituent functions are chosen as
\begin{equation}
\begin{array}{l}
F_1 = \sum_{i=1}^{p} (x_{u1}^{i})^2,\\
F_2 = - \sum_{i=1}^{q} (x_{l1}^{i})^2,\\
F_3 = \sum_{i=1}^{r} (x_{u2}^{i})^2 - \sum_{i=1}^{r} (x_{u2}^{i} - \log x_{l2}^{i})^2,\\
f_1 = \sum_{i=1}^{p} (x_{u1}^{i})^2,\\
f_2 = \sum_{i=1}^{q} (x_{l1}^{i})^2,\\
f_3 = \sum_{i=1}^{r} (x_{u2}^{i} - \log x_{l2}^{i})^2.\\
\end{array}
\end{equation}
The upper and lower level constraints are as follows:
\begin{equation}
\begin{array}{l}
\mbox{Upper level constraints}\\
G_j: x_{u2}^{j} \ge \frac{1}{\sqrt{r}} + \log x_{l2}^{j}, \hspace{1mm} \forall \hspace{1mm} j \in \{1,2,\ldots,r\},\\
\mbox{Lower level constraint}\\
g_1: \sum_{i=1}^{r} (x_{u2}^{i} - \log x_{l2}^{i})^2 \ge 1.\\
\end{array}
\end{equation}
The range of variables is as follows:
\begin{equation}
\begin{array}{l}
x_{u1}^{i} \in [-5,10], \hspace{2mm} \forall \hspace{2mm} i \in \{1,2,\ldots,p\},\\
x_{u2}^{i} \in [-1,1], \hspace{2mm} \forall \hspace{2mm} i \in \{1,2,\ldots,r\},\\
x_{l1}^{i} \in [-5,10], \hspace{2mm} \forall \hspace{2mm} i \in \{1,2,\ldots,q\},\\
x_{l2}^{i} \in [\frac{1}{e},e], \hspace{2mm} \forall \hspace{2mm} i \in \{1,2,\ldots,r\}.
\end{array}
\end{equation}
Relationship between upper level variables and lower level optimal variables is given as follows:
\begin{equation}
\begin{array}{l}
x_{l1}^{i} = 0, \hspace{2mm} \forall \hspace{2mm} i \in \{1,2,\ldots,q\},\\
\boldx_{l2} : \sum_{i=1}^{r} (x_{u2}^{i} - \log x_{l2}^{i})^2 = 1.
\end{array}
\end{equation}
The values of the variables at the optima are $\boldx_{u1}=0$, $\boldx_{u2}=0$, $\boldx_{l1}=0$, and $\boldx_{l2}=\log^{-1}\frac{-1}{\sqrt{r}}$. The upper level function value is $-1$ and the
lower level function value is $+1$ at the optima.

\begin{figure}
\begin{center}
\epsfig{file=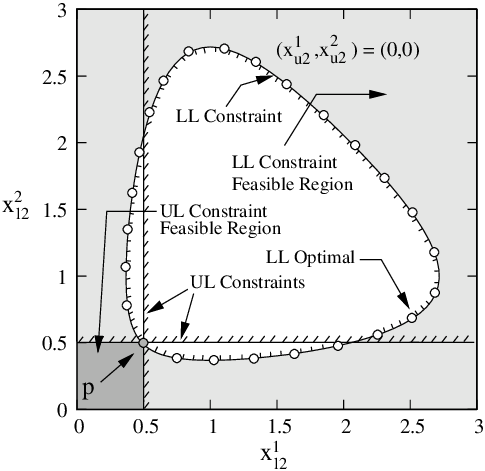,width=0.45\linewidth}
\caption{Feasible and infeasible regions of SMD11 for a particular upper level vector.}
\label{fig:smd11}
\end{center}
\end{figure}

Figure \ref{fig:smd11} shows the constraints at the upper as well as the lower level when $r=2$. In this example, there is one constraint at the lower level and two constraints at the upper level. All the solutions on the lower level constraint represent optimal solutions to the lower level $f_3$. When $\boldx_{l1}=0$, such that the function $f_2$ is also optimal, the solutions on the constraint are optimal solutions to the lower level problem  for a given $\boldx_u$. It can be observed that the two constraints at the upper level eliminate all the lower level optimal solutions except one. The figure shows the feasible region with respect to upper level constraints for the upper level problem. However, only point $\mathbf{p}$ represents a feasible solution for the upper level problem for a given $\boldx_u$, as it is the lower level optimal solution lying in the upper level constraint feasible region. This problem differs from SMD6, which also contained multiple global solutions at the lower level, in two ways. First, multiple global solutions at the lower level are introduced by lower level constraints in this problem, whereas in the previous problem it was the lower level objective function that was entirely responsible for introducing multiple global solutions. Second, out of the multiple global solutions from the lower level, a single solution is selected based on upper level constraints, whereas in the previous problem all the lower level global solutions were feasible but only one of those solutions had the best upper level objective value.

\subsection{SMD12}
This test problem is a combination of the previous two test problems, and involves a number of difficulties. The test problem has scalable constraints at both levels, and the constraints are functions of both upper as well as lower level variables. At the same time, any lower level optimization problem for a given set of upper level variables has multiple global optima. All the lower level constraints are active at the bilevel optimum. The constituent functions are chosen as
\begin{equation}
\begin{array}{l}
F_1 = \sum_{i=1}^{p} (x_{u1}^{i} - 2)^2,\\
F_2 = \sum_{i=1}^{q} (x_{l1}^{i})^2,\\
F_3 = \sum_{i=1}^{r} (x_{u2}^{i} - 2)^2 + \sum_{i=1}^{r} \tan |x_{l2}^{i}| - \sum_{i=1}^{r} (x_{u2}^{i} - \tan x_{l2}^{i})^2,\\
f_1 = \sum_{i=1}^{p} (x_{u1}^{i})^2,\\
f_2 = \sum_{i=1}^{q} (x_{l1}^{i} - 2)^2,\\
f_3 = \sum_{i=1}^{r} (x_{u2}^{i} - \tan x_{l2}^{i})^2.\\
\end{array}
\end{equation}
The upper and lower level constraints are as follows:
\begin{equation}
\begin{array}{l}
\mbox{Upper level constraints}\\
x_{u2}^{i} - \tan x_{l2}^{i} \ge 0, \hspace{1mm} \forall \hspace{1mm} i \in \{1,2,\ldots,r\},\\
x_{u1}^{j} - \sum_{i=1,i \ne j}^{p} (x_{u1}^{i})^3 - \sum_{i=1}^{r} (x_{u2}^{i})^3  \ge 0, \hspace{1mm} \forall \hspace{1mm} j \in \{1,2,\ldots,p\},\\
x_{u2}^{j} - \sum_{i=1,i \ne j}^{r} (x_{u2}^{i})^3 - \sum_{i=1}^{p} (x_{u1}^{i})^3  \ge 0, \hspace{1mm} \forall \hspace{1mm} j \in \{1,2,\ldots,r\},\\
\mbox{Lower level constraints}\\
\sum_{i=1}^{r} (x_{u2}^{i} - \tan x_{l2}^{i})^2 \ge 1,\\
x_{l1}^{j} - \sum_{i=1,i \ne j}^{p} (x_{l1}^{i})^3, \hspace{1mm} \forall \hspace{1mm} j \in \{1,2,\ldots,q\}.\\
\end{array}
\end{equation}
The range of variables is as follows:
\begin{equation}
\begin{array}{l}
x_{u1}^{i} \in [-5,10], \hspace{2mm} \forall \hspace{2mm} i \in \{1,2,\ldots,p\},\\
x_{u2}^{i} \in [-14.10,14.10], \hspace{2mm} \forall \hspace{2mm} i \in \{1,2,\ldots,r\},\\
x_{l1}^{i} \in [-5,10], \hspace{2mm} \forall \hspace{2mm} i \in \{1,2,\ldots,q\},\\
x_{l2}^{i} \in (-1.5,1.5), \hspace{2mm} \forall \hspace{2mm} i \in \{1,2,\ldots,r\}.
\end{array}
\end{equation}
Relationship between upper level variables and lower level optimal variables is given as follows:
\begin{equation}
\begin{array}{l}
x_{l1}^{i} = \frac{1}{\sqrt{q-1}}, \hspace{2mm} \forall \hspace{2mm} i \in \{1,2,\ldots,q\},\\
\boldx_{l2} : \sum_{i=1}^{r} (x_{u2}^{i} - \tan x_{l2}^{i})^2 = 1.
\end{array}
\end{equation}
The values of the variables at the optima are $\boldx_{u1}=\frac{1}{\sqrt{p+r-1}}$, $\boldx_{u2}=\frac{1}{\sqrt{p+r-1}}$, $\boldx_{l1}=\frac{1}{\sqrt{q-1}}$, and $\boldx_{l2}=\tan^{-1}(\frac{1}{\sqrt{p+r-1}}-\frac{1}{\sqrt{r}})$. 

\subsection{Summary and Precautions}
The properties of the SMD test problems are summarized in Table \ref{tab:summary}. In the table, $N$ denotes ${\rm No}$ and $Y$ denotes ${\rm Yes}$. It can be observed that the 12 test problems are a good mix of various difficulties that we discussed in the prior sections. We have tried to put the problems in an increasing order of difficulty. The last test problem can be observed to contain most of the difficulties except multi-modalities. This table will be helpful in testing algorithms for bilevel optimization. For example, if a new algorithm is able to solve SMD1 but not SMD2, one readily concludes that the algorithm is unable to handle a conflict. Similarly, if the algorithm is able to solve SMD1 and SMD2 but not SMD3 and SMD4, one would infer that the algorithm is unable to handle lower level multi-modality. Such information will be useful for an algorithm developer, as it helps him to identify the specific weaknesses in his approach, which he needs to improve on.

Authors would like to caution the developers against heavily relying on test problems alone to draw conclusions about the performance of the algorithm. The test problems are useful at the initial stages of algorithm development to evaluate the performance of an algorithm across various difficulty frontiers. 
However, it might not always be possible for a test-suite to provide difficulties that can be offered by complex real-world problems. Therefore, it is very important to note that the suggested test problems are not a replacement for realistic problems. It is important for researchers to focus on real-world problems as well along with the test suites to evaluate their procedures.

In the field of evolutionary multi-objective optimization, the test-suites have been quite famous and the developers are often found to draw strong conclusions based on the performance of the algorithms on these test-suites. One of the caveats is to exploit the structured nature of these test-suites to report better performance for their approaches. For example, in the proposed test-suite many of the test problems contain variable separable functions. These test problems would certainly be relatively easier to solve if an algorithm exploits this property of the test problems. Such algorithms would deteriorate drastically if these functions are rotated by multiplying the variables with a transformation matrix. On the other hand, an algorithm that does not exploit this property will be indifferent between the variable separable and the rotated test problems. It is important to utilize this knowledge about the test problems rather constructively to evaluate the extent to which an algorithm is exploiting the variable separability of the test problems. The authors would like the users to be careful about knowingly or unknowingly exploiting any such structure of the proposed test problems.

\begin{table*}[!ht]
\begin{center}
\vspace{0mm}
\caption{Properties of SMD test problems.} 
\label{tab:summary}
\epsfig{file=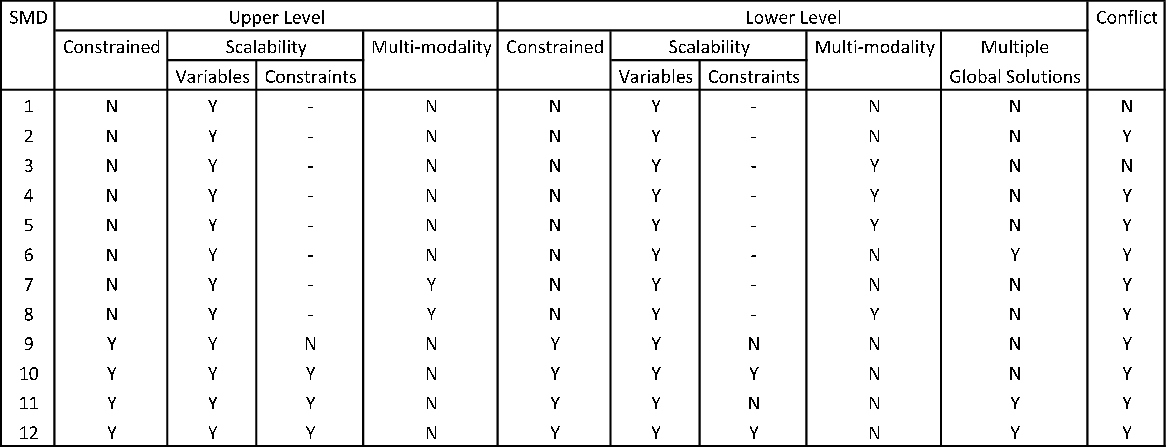,width=0.95\linewidth}
\vspace{0mm}
\end{center}
\end{table*}

\section{Baseline Solution Methodology}\label{sec:nested-algorithm}
In this section, we describe the solution methodology used to solve the constructed test problems. The suggested procedure is a nested bilevel evolutionary algorithm, and requires that a lower level optimization task be solved for every new set of upper level variables produced using the genetic operators. The method relies on a steady state single objective real coded genetic algorithm to solve the problems at both levels. We have implemented a modified version of the procedures \citep{my-cec05,my-cec06} with two levels, which is used to handle the bilevel test problems. 
A step-by-step procedure for the algorithm is described as follows:

\subsection{Upper Level Optimization Procedure}\label{sec:upperLevel}
\textit{}
\vspace{-7mm}

\textit{Step 1: Initialization Scheme.} Initialize a random population ($N_p$) of upper level variables. For each upper level population member execute a lower level optimization procedure to determine the corresponding optimal lower level variables. Assign upper level fitness based on the upper level function value and constraints.

\textit{Step 2: Selection of upper level parents.} Choose $2\mu$ population members from the previous population and conduct a tournament selection to determine $\mu$ parents.

\textit{Step 3: Evolution at the upper level.} Perform a crossover (Refer to Subsection \ref{sec:crossover}) and a polynomial mutation to create $\lambda$ offspring. This provides the upper level variables for each offspring.

\textit{Step 4: Lower level optimization.} Solve the lower level optimization problem (Refer to Subsection \ref{sec:lowerLevel}) for each offspring. This provides the lower level variables for each offspring.

\textit{Step 5: Evaluate offspring.} Combine the upper level variables with the corresponding optimal lower level variables for each offspring. Evaluate all the offspring based on upper level function value and constraints.

\textit{Step 6: Population update.} Choose $r$ random members from the parent population and pool them with the $\lambda$ offspring. The best $r$ members from the pool replace the chosen $r$ members from the population.

\textit{Step 7: Termination check.} Proceed to the next generation (Step 2) if the termination check (Refer to Subsection \ref{sec:termination}) is false.

\subsection{Lower Level Optimization Procedure}\label{sec:lowerLevel}
The lower level optimization procedure is similar to the upper level procedure except the initialization step which differs slightly. In the following, we provide the steps involved during the lower level optimization task. Let the lower level population size be $n_p$, and the upper level member being optimized be $\boldx_{u}^{0}$.

\textit{}
\vspace{-7mm}

\textit{Step 1:} If the execution is transferred from Step 1 of the upper level optimization task then go to (a) otherwise go to (b),

\quad \textit{a:} Initialize $n_p$ lower level member randomly, and assign lower level fitness based on the lower level function value and constraints. Go to Step 2.

\quad \textit{b:} Initialize $n_p-1$ lower level members randomly. Determine the member closest to $\boldx_{u}^{0}$ in the upper level population. The lower level optimal variables from the closest upper level member becomes the $n_{p}^{th}$ member in the lower level population. Assign lower level fitness based on the lower level function value and constraints. Go to Step 2.

\textit{Step 2:} Choose $2\mu$ members randomly from the lower level population. Perform a tournament selection with respect to lower level fitness to generate $\mu$ parents. 

\textit{Step 3:} Perform crossover and mutation to generate $\lambda$ offspring.

\textit{Step 4:} Evaluate each offspring with respect to lower level function and constraints.

\textit{Step 5:} Choose $r$ members randomly from the lower level population and pool them with the $\lambda$ lower level offspring. The best $r$ members with respect to lower level fitness replace the chosen $r$ members from the lower level population.

\textit{Step 6:} Proceed to the next generation (Step 2) if the termination check (Refer to Subsection \ref{sec:termination}) is false.

\subsection{Parameters}
The parameters in the algorithm were fixed as $\mu=3$, $\lambda=3$ and $r=2$. Probability of crossover was fixed as $0.9$ and the probability of mutation was fixed as $0.1$. The crossover operator requires two parameters $\omega_{\xi}$ and $\omega_{\eta}$, which are fixed as suggested in the next subsection.

\subsection{Crossover Operator}
\label{sec:crossover}
The crossover operator used at both levels is similar to the PCX operator proposed in \cite{my-cec06} with minor modifications. The operator creates an offspring from three parents, when one of the three parents is chosen as the index parent as follows,
\begin{equation}
\mathbf{c} = \mathbf{x_p} + \omega_{\xi}\mathbf{d} + \omega_{\eta}\frac{\mathbf{p_2}-\mathbf{p_1}}{2}.
\label{eq:child}
\end{equation}
The terms used in the above equation are defined as,
\begin{itemize}\itemsep1pt \parskip0pt \parsep0pt
	\item $\mathbf{x_p}$ is the {\em index\/} parent
	\item $\mathbf{d}=\mathbf{x_{p}}-\mathbf{w}$, where $\mathbf{w}$ is the mean of $\mu$ parents
	\item $\mathbf{p_1}$ and $\mathbf{p_2}$ are the other two parents
	\item $\omega_{\xi}=0.1$ and $\omega_{\eta}=\sum_{i=1}^{m_v} \frac{m_v}{|x_{p}^{i}-w^{i}|}$ are the two parameters, where $v \in \{u,l\}$ such that $m_u$ is the number of variables at the upper level and $m_l$ is the number of variables at the lower level.
\end{itemize}
The two parameters $\omega_{\xi}$ and $\omega_{\eta}$, describe the extent of variations along the respective directions. While creating $\lambda=3$ offspring from $\mu=3$ parents, each parent is chosen as an index parent at a time.

\subsection{Constraint Handling}
We define the constraint violation as the sum of violations of all the constraints at the respective levels. If a member at a particular level has a smaller constraint violation, then it is always preferred over a member with a higher constraint violation at the same level. A member with no constraint violation is deemed to be feasible, and is considered better than any of the other infeasible members. While comparing two feasible members, the member with a smaller function value at the level is preferred.

\subsection{Termination Check}
\label{sec:termination}
The algorithm uses a variance based termination criteria at both levels. When the value of $\alpha_j$, described in the following equation becomes less than $\alpha_{stop}$, the optimization task terminates. In the following, we state the termination criteria at the lower level, which can be similarly extended to the upper level. Let the variance of the lower level population members at generation $j$ for each lower level variable $i$ be $v_{j}^{i}$. If the number of lower level variables is $m_l$, then $\alpha$ is computed as,
\begin{equation}
\begin{array}{l}
	\alpha_j = \sum_{i=1}^{m_l} \frac{v_{j}^i}{v_{0}^i}.
\label{eq:criteria}
\end{array}
\end{equation}
The value of $\alpha_j$ usually lies between 0 and 1 in Equation~\ref{eq:criteria}. In the above equation, $v_{0}^i$ denotes the variance for the variable $i$ in the initial lower level population. For the lower level, the value of $\alpha_{stop}$ is set as $10^{-5}$, and for the upper level the value of $\alpha_{stop}$ is set as $10^{-4}$.

\section{Results}\label{sec:results}
In this section, we provide the results obtained from solving the proposed test problems using the bilevel evolutionary algorithm. The described nested bilevel evolutionary algorithm is a naive scheme, and any intelligent bilevel approach should be expected to produce better results with lesser computational expense. The results are intended as benchmark, and the performance of other schemes may be compared in terms of percentage saving obtained when compared to the proposed nested scheme. We performed $11$ runs for each of the test problems with $5$, $10$ and $20$ dimensions. In case of 5 dimensions, for SMD1 to SMD5 and SMD7 to SMD12 we choose $p=1$, $q=2$ and $r=1$, and for SMD6 we choose $p=1$, $q=0$, $r=1$ and $s=2$. In case of 10 dimensions, for SMD1 to SMD5 and SMD7 to SMD12 we choose $p=3$, $q=3$ and $r=2$, and for SMD6 we choose $p=3$, $q=1$, $r=2$ and $s=2$. The upper level population size $N_p$ and the lower level population size $n_p$ were chosen as 30 for the $5$ dimensional case. Both population sizes were chosen as 50 and 100 for 10 and 20 dimensional cases respectively.

Results for $5$ dimensional test problems are reported in Tables~\ref{tab:table1a} and \ref{tab:table1b}. Table~\ref{tab:table1a} provides the best, median, and worst number of function evaluations at upper and lower levels. The accuracy achieved and the number of times lower level optimization was performed in a single execution of the bilevel optimization run are reported in Table~\ref{tab:table1b}. Similar results for $10$ dimensional test problems are reported in Tables~\ref{tab:table2a} and \ref{tab:table2b}. For $20$ dimensional test problems, we report only the best, median, and worst function evaluations in Table~\ref{tab:table3a}.

\begin{table*}[hbt]
\caption{Function evaluations (FE) for the upper level (UL) and the lower level (LL) from 11
  runs for 5 dimensional test problems.} 
\label{tab:table1a}
{\small\begin{center}
\begin{tabular}{|c|c|c|c|c|c|c|} \hline
Pr. No.	&	\multicolumn{2}{|c|}{Best}	&	\multicolumn{2}{|c|}{Median}	&	\multicolumn{2}{|c|}{Worst}	\\	\cline{2-7}
	&		\multicolumn{1}{|c|}{Total LL}	&	\multicolumn{1}{|c|}{Total UL}	&	\multicolumn{1}{|c|}{Total LL}	&	\multicolumn{1}{|c|}{Total UL}	&	\multicolumn{1}{|c|}{Total LL}	&\multicolumn{1}{|c|}{Total UL}	\\	
	&	\multicolumn{1}{|c|}{FE} 	&
        \multicolumn{1}{|c|}{FE}	&\multicolumn{1}{|c|}{FE}         &\multicolumn{1}{|c|}{FE}        &
        \multicolumn{1}{|c|}{FE}& \multicolumn{1}{|c|}{FE}	\\ \hline	
SMD1	&	256858	&	438	&	375488	&	668	&	582770	&	1008	\\	\hline
SMD2	&	196744	&	380	&	332197	&	628	&	613221	&	1102	\\	\hline
SMD3	&	262703	&	488	&	315598	&	604	&	439316	&	844	\\	\hline
SMD4	&	259486	&	420	&	366294	&	608	&	480675	&	796	\\	\hline
SMD5	&	222078	&	444	&	457265	&	930	&	610108	&	1232	\\	\hline
SMD6	&	334763	&	540	&	427114	&	696	&	585358	&	936	\\	\hline
SMD7	&	246375	&	468	&	333629	&	652	&	685029	&	1342	\\	\hline
SMD8	&	443430	&	812	&	582583	&	1008	&	1218196	&	2076	\\	\hline
SMD9	&	183231	&	330	&	284648	&	514	&	395735	&	696	\\	\hline
SMD10	&	179986	&	480	&	277696	&	758	&	501639	&	1316	\\	\hline
SMD11	&	11489609	&	4348	&	13408524	&	5086	&	20540610	&	7764	\\	\hline
SMD12	&	6211173	&	354	&	12950512	&	738	&	20983708	&	1196	\\	\hline
\end{tabular}
\end{center}}
\end{table*}

\begin{table*}[hbt]
\caption{Accuracy for the upper and lower levels, and the lower level calls from 11
  runs for 5 dimensional test problems.} 
\label{tab:table1b}
\begin{center}
\begin{tabular}{|c|c|c|c|c|} \hline
Pr. No. & Median & Median & Median &  \\ \cline{2-4}
	& UL Accuracy & LL Accuracy & LL Calls &  $\frac{\mbox{LL Evals}}{\mbox{LL Calls}}$ \\ \hline
SMD1 & 0.000114 & 0.000087 & 668 & 563.89 	\\	\hline
SMD2 & 0.000073 & 0.000016 & 628 & 533.01 	\\	\hline
SMD3 & 0.000054 & 0.000055 & 604 & 536.47 	\\	\hline
SMD4 & 0.000023 & 0.000057 & 608 & 607.80 	\\	\hline
SMD5 & 0.000002 & 0.000009 & 930 & 507.82 	\\	\hline
SMD6 & 0.000108 & 0.000061 & 696 & 604.64 	\\	\hline
SMD7 & 0.000016 & 0.000177 & 652 & 533.84 	\\	\hline
SMD8 & 0.000174 & 0.000027 & 1008 & 562.69 	\\	\hline
SMD9 & 0.000017 & 0.000054 & 514 & 553.54 	\\	\hline
SMD10 & 0.034759 & 0.018510 & 758 & 367.04 	\\	\hline
SMD11 & 0.0131643 & 0.0129893 & 5086 & 2635.64 \\	\hline
SMD12 & 0.032372 & 0.000206 & 738 & 19202.32 \\	\hline
\end{tabular}
\end{center}
\end{table*}

\begin{table*}[hbt]
\caption{Function evaluations (FE) for the upper level (UL) and the lower level (LL) from 11
  runs for 10 dimensional test problems. A `x' denotes that the algorithm terminated far away ($\Delta F \ge 0.1$) from the optimal solution. A `-' denotes that a feasible solution could not be obtained for the test problem.} 
\label{tab:table2a}
{\small\begin{center}
\begin{tabular}{|c|c|c|c|c|c|c|} \hline
Pr. No.	&	\multicolumn{2}{|c|}{Best}	&	\multicolumn{2}{|c|}{Median}	&	\multicolumn{2}{|c|}{Worst}	\\	\cline{2-7}
	&		\multicolumn{1}{|c|}{Total LL}	&	\multicolumn{1}{|c|}{Total UL}	&	\multicolumn{1}{|c|}{Total LL}	&	\multicolumn{1}{|c|}{Total UL}	&	\multicolumn{1}{|c|}{Total LL}	&\multicolumn{1}{|c|}{Total UL}	\\	
	&	\multicolumn{1}{|c|}{FE} 	&
        \multicolumn{1}{|c|}{FE}	&\multicolumn{1}{|c|}{FE}         &\multicolumn{1}{|c|}{FE}        &
        \multicolumn{1}{|c|}{FE}& \multicolumn{1}{|c|}{FE}	\\ \hline	
SMD1 & 862653 & 1080 & 1623356 & 2534 & 2074334 & 3488	\\	\hline
SMD2 & 1055976 & 1398 & 1467246 & 2366 & 2114442 & 3418	\\	\hline
SMD3 & 900358 & 1210 & 1383632 & 2278 & 1805562 & 2862	\\	\hline
SMD4 & 566344 & 678 & 1087632 & 1598 & 1314986 & 2028	\\	\hline
SMD5 & 1226344 & 1620 & 1993124 & 2890 & 2483442 & 3492	\\	\hline
SMD6 & 1225742 & 1502 & 2224450 & 2936 & 3786498 (x) & 4278 (x)	\\	\hline
SMD7 & 932460 & 1382 & 1566481 & 2394 & 2435994 (x) & 3858(x) \\	\hline
SMD8 & 1457480 & 2116 & 2710132 & 4188 & 5294734 (x) & 5986 (x)  \\	\hline
SMD9 & - & - & - & - & - & -  \\	\hline
SMD10 & - & - & - & - & - & -  \\	\hline
SMD11 & -& -& -& -& -& - \\ \hline
SMD12 & -& -& -& -& -& - \\ \hline
\end{tabular}
\end{center}}
\end{table*}

\begin{table*}[hbt]
\caption{Accuracy for the upper and lower levels, and the lower level calls from 11
  runs for 10 dimensional test problems. A `-' denotes that a feasible solution could not be obtained for the test problem.} 
\label{tab:table2b}
\begin{center}
\begin{tabular}{|c|c|c|c|c|} \hline
Pr. No. & Median & Median & Median &  \\ \cline{2-4}
	& UL Accuracy & LL Accuracy & LL Calls &  $\frac{\mbox{LL Evals}}{\mbox{LL Calls}}$ \\ \hline
SMD1 & 0.000332 & 0.000018 & 2534 & 644.54	\\	\hline
SMD2 & 0.000066 & 0.000011 & 2366 & 653.36	\\	\hline
SMD3 & 0.000359 & 0.000033 & 2278 & 655.76	\\	\hline
SMD4 & 0.000286 & 0.000027 & 1598 & 685.43	\\	\hline
SMD5 & 0.000052 & 0.000009 & 2890 & 716.82	\\	\hline
SMD6 & 0.001435 & 0.000082 & 2936 & 768.34	\\	\hline
SMD7 & 0.006263 & 0.000127 & 2394 &	654.34\\	\hline
SMD8 & 0.003122 & 0.000157 & 4188 &	647.12\\	\hline
SMD9 & - & - & - & - \\	\hline
SMD10 & - & - & - & - 	\\	\hline
SMD11 & - & - & - & - 	\\	\hline
SMD12 & - & - & - & - 	\\	\hline
\end{tabular}
\end{center}
\end{table*}

\begin{table*}[!ht]
\caption{Function evaluations (FE) for the upper level (UL) and the lower level (LL) from 11
  runs for 20 dimensional test problems. A `x' denotes that the algorithm terminated far away ($\Delta F \ge 0.1$) from the optimal solution. A `-' denotes that a feasible solution could not be obtained for the test problem.} 
\label{tab:table3a}
{\small\begin{center}
\begin{tabular}{|c|c|c|c|c|c|c|} \hline
Pr. No.	&	\multicolumn{2}{|c|}{Best}	&	\multicolumn{2}{|c|}{Median}	&	\multicolumn{2}{|c|}{Worst}	\\	\cline{2-7}
	&		\multicolumn{1}{|c|}{Total LL}	&	\multicolumn{1}{|c|}{Total UL}	&	\multicolumn{1}{|c|}{Total LL}	&	\multicolumn{1}{|c|}{Total UL}	&	\multicolumn{1}{|c|}{Total LL}	&\multicolumn{1}{|c|}{Total UL}	\\	
	&	\multicolumn{1}{|c|}{FE} 	&
        \multicolumn{1}{|c|}{FE}	&\multicolumn{1}{|c|}{FE}         &\multicolumn{1}{|c|}{FE}        &
        \multicolumn{1}{|c|}{FE}& \multicolumn{1}{|c|}{FE}	\\ \hline	
SMD1 & 3105178 & 3210 & 5262456 & 5248 & 6868944 & 7378 \\	\hline
SMD2 & 2166384 & 3326 & 4102678 & 4052 & 5803812 & 7076 \\	\hline
SMD3 & 3696032 & 3220 & 4814112 & 4282 & 7015724 & 6314 \\	\hline
SMD4 & 2017734 & 2454 & 2755534 & 3110 & 4364876 & 5296 \\	\hline
SMD5 & 4574482 & 4488 & 8800232 & 7004 & 12064566 & 9290 \\	\hline
SMD6 & 5026522 & 3530 & 8448154 & 6962 & 12448922 (x) & 9978 (x) \\	\hline
SMD7 & - & - & - & - & - & -  \\	\hline
SMD8 & - & - & - & - & - & -  \\	\hline
SMD9 & - & - & - & - & - & -  \\	\hline
SMD10 & - & - & - & - & - & -  \\	\hline
SMD11 & -& -& -& -& -& - \\ \hline
SMD12 & -& -& -& -& -& - \\ \hline
\end{tabular}
\end{center}}
\end{table*}

The nested bilevel evolutionary algorithm was able to solve all the test problems with $5$ dimensions. We consider a test problem solved if the difference between the function value achieved by the algorithm and the optimal function value is no more than $0.1$. However, the number of function evaluations required to obtain the optimal solutions in each of the test problems is large. The function evaluations at the upper level are much smaller, as compared to the function evaluations at the lower level. A large number of lower level function evaluations are required, as a lower level optimization task is executed for each upper level vector. For every newly created upper level vector, we first find the lower level optimal solution and then evaluate the upper level function value. Therefore, the number of function evaluations at the upper level is same as the number of times the lower level optimization task is executed.
When the size of the test problems is increased to $10$, we observe that the number of function evaluations increase significantly. The nested approach is able to successfully solve the first 5 test problems in all the runs. For test problems SMD6, SMD7 and SMD8, it is unable to solve the problems in all the runs, rather it arrives at the optimal solutions for more than $50\%$ of the runs. For SMD6 the success rate was $82\%$, for SMD7 it was $73\%$ and for SMD8 it was $63\%$. The nested approach fails to handle the constrained test problems for the chosen algorithm parameters. The lower level problems could not be completely solved for SMD9 to SMD12, which introduced infeasible members at the upper level. In case of $20$ dimensional test problems, the nested approach is able to solve SMD1 to SMD5 for all the runs. It is able to handle SMD6 in $63\%$ of the runs, but fails to handle SMD7 to SMD12.

The results demonstrate that a high number of function evaluations are required to solve bilevel problems. With an increase in the number of dimensions, the complexity increases significantly and the available computational resources quickly become insufficient to solve larger versions of the problems. In this paper, we utilize a global optimizer at both levels, which successfully solved smaller versions of the test problems, but failed for constrained test problems with high dimensions. Given, the complex nature of bilevel optimization problems, evolutionary algorithms might be a useful approach to follow. However, using evolutionary algorithms alone would demand a large number of function evaluations to solve even simple bilevel problems. Therefore, an intelligent approach which utilizes results from the classical literature within an evolutionary algorithm might be a feasible direction towards handling such problems. The set of test problems proposed in this paper would be useful to evaluate such algorithms across various difficulties which a bilevel optimization problem could offer.

\section{Conclusions}
In this paper we have provided a test problem construction procedure for unconstrained as well as constrained bilevel optimization. The procedure offers the flexibility to control the difficulties at the two levels individually as well as collectively. To demonstrate the framework, we have created a test-bed of 12 bilevel optimization problems, out of which 8 are unconstrained and 4 are constrained. The test-suite contains problems, which are scalable in terms of number of variables as well as constraints. Moreover, the optimal solutions for all the test problems are clearly identified, which would be useful in testing and evaluating bilevel optimization algorithms. The test problem construction procedure should allow researchers to create additional test problems by varying the basic functions used in different test problems. As a benchmark for comparison, we have provided results from a nested bilevel evolutionary scheme, which utilizes a global optimizer at both levels. Five and ten-variable instances of all the test problems have been solved, which demonstrate the high computational requirement of bilevel problems even for smaller instances. This amply indicates that the solution of bilevel problems, even with an evolutionary algorithm, is a challenging task and more attention must be devoted to develop computationally faster algorithms.

\section*{Acknowledgments}
Authors A. Sinha and P. Malo wish to thank the Wallenberg foundation and Liikesivistysrahasto for supporting this study. P. Malo acknowledges the support provided by the Emil Aaltonen foundation. K. Deb acknowledges start-up grant from Department of Electrical and Computer Engineering and College of Engineering at Michigan State University, East Lansing, USA.

\small


\begin{thebibliography}{}

\bibitem[Aiyoshi and Shimizu, 1981]{aiyoshi81}
Aiyoshi, E. and Shimizu, K. (1981).
\newblock Hierarchical decentralized systems and its new solution by a barrier
  method.
\newblock {\em IEEE Transactions on Systems, Man, and Cybernetics},
  11:444--449.

\bibitem[Bard and Falk, 1982]{bard82}
Bard, J. and Falk, J. (1982).
\newblock An explicit solution to the multi-level programming problem.
\newblock {\em Computers and Operations Research}, 9:77--100.

\bibitem[Bard, 1983]{bard83}
Bard, J.~F. (1983).
\newblock Coordination of multi-divisional firm through two levels of
  management.
\newblock {\em Omega}, 11(5):457--465.

\bibitem[Bianco et~al., 2009]{bianco-kkt}
Bianco, L., Caramia, M., and Giordani, S. (2009).
\newblock A bilevel flow model for hazmat transportation network design.
\newblock {\em Transportation Research Part C: Emerging technologies},
  17(2):175--196.

\bibitem[Brotcorne et~al., 2001]{brotcorne01}
Brotcorne, L., Labbe, M., Marcotte, P., and Savard, G. (2001).
\newblock A bilevel model for toll optimization on a multicommodity
  transportation network.
\newblock {\em Transportation Science}, 35(4):345--358.

\bibitem[Calamai and Vicente, 1992]{calamai92}
Calamai, P.~H. and Vicente, L.~N. (1992).
\newblock Generating linear and linear-quadratic bilevel programming problems.
\newblock {\em SIAM J. Sci. Comput.}, 14(1):770--782.

\bibitem[Calamai and Vicente, 1994]{calamai94}
Calamai, P.~H. and Vicente, L.~N. (1994).
\newblock Generating quadratic bilevel programming test problems.
\newblock {\em ACM Trans. Math. Software}, 20(1):103--119.

\bibitem[Colson et~al., 2007]{colson}
Colson, B., Marcotte, P., and Savard, G. (2007).
\newblock An overview of bilevel optimization.
\newblock {\em Annals of Operational Research}, 153:235--256.

\bibitem[Constantin and Florian, 1995]{constantin95}
Constantin, I. and Florian, M. (1995).
\newblock Optimizing frequencies in a transit network: a nonlinear bi-level
  programming approach.
\newblock {\em International Transactions in Operational Research}, 2(2):149 --
  164.

\bibitem[Deb et~al., 2002]{pcx}
Deb, K., Anand, A., and Joshi, D. (2002).
\newblock A computationally efficient evolutionary algorithm for real-parameter
  optimization.
\newblock {\em Evolutionary Computation Journal}, 10(4):371--395.

\bibitem[Deb and Sinha, 2010]{my-ecj10}
Deb, K. and Sinha, A. (2010).
\newblock An efficient and accurate solution methodology for bilevel
  multi-objective programming problems using a hybrid evolutionary-local-search
  algorithm.
\newblock {\em Evolutionary Computation Journal}, 18(3):403--449.

\bibitem[Dempe, 2002]{dempe02}
Dempe, S. (2002).
\newblock {\em {Foundations of Bilevel Programming}}.
\newblock Kluwer Academic Publishers, Secaucus, NJ, USA.

\bibitem[Dempe et~al., 2006]{dempe-dutta}
Dempe, S., Dutta, J., and Lohse, S. (2006).
\newblock Optimality conditions for bilevel programming problems.
\newblock {\em Optimization}, 55(5--6):505--524.

\bibitem[Frantsev et~al., 2012]{my-ifac12}
Frantsev, A., Sinha, A., and Malo, P. (2012).
\newblock Finding optimal strategies in multi-period stackelberg games using an
  evolutionary framework.
\newblock In {\em IFAC Workshop on Control Applications of Optimization
  (IFAC-2009)}. Elsevier.

\bibitem[Fudenberg and Tirole, 1993]{fudenberg93}
Fudenberg, D. and Tirole, J. (1993).
\newblock {\em Game theory}.
\newblock MIT Press.

\bibitem[Herskovits et~al., 2000]{bilevel-KKT1}
Herskovits, J., Leontiev, A., Dias, G., and Santos, G. (2000).
\newblock Contact shape optimization: A bilevel programming approach.
\newblock {\em Struct Multidisc Optimization}, 20:214--221.

\bibitem[Kirjner-Neto et~al., 1998]{kirjnerneto98}
Kirjner-Neto, C., Polak, E., and Kiureghian, A. (1998).
\newblock An outer approximations approach to reliability-based optimal design
  of structures.
\newblock {\em Journal of Optimization Theory and Applications}, 98(1):1--16.

\bibitem[Migdalas, 1995]{migdalas95}
Migdalas, A. (1995).
\newblock Bilevel programming in traffic planning: Models, methods and
  challenge.
\newblock {\em Journal of Global Optimization}, 7(4):381--405.

\bibitem[Mitsos and Barton, 2006]{mitsos06}
Mitsos, A. and Barton, P.~I. (2006).
\newblock A test set for bilevel programs.
\newblock http://yoric.mit.edu/download/Reports/bileveltestset.pdf.

\bibitem[Moshirvaziri et~al., 1996]{moshirvaziri96}
Moshirvaziri, K., Amouzegar, M.~A., and Jacobsen, S.~E. (1996).
\newblock Test problem construction for linear bilevel programming problems.
\newblock {\em Journal of Global Optimization}, 8(3):235--243.

\bibitem[Sinha et~al., 2012]{my-cec12a}
Sinha, A., Malo, P., and Deb, K. (2012).
\newblock Unconstrained scalable test problems for single-objective bilevel
  optimization.
\newblock In {\em 2012 IEEE Congress on Evolutionary Computation (CEC-2012)}.
  IEEE Press.

\bibitem[Sinha et~al., 2013]{my-cec13}
Sinha, A., Malo, P., Frantsev, A., and Deb, K. (2013).
\newblock Multi-objective stackelberg game between a regulating authority and a
  mining company: A case study in environmental economics.
\newblock In {\em 2013 IEEE Congress on Evolutionary Computation (CEC-2013)}.
  IEEE Press.

\bibitem[Sinha et~al., 2014]{my-caor13}
Sinha, A., Malo, P., Frantsev, A., and Deb, K. (2014).
\newblock Finding optimal strategies in a multi-period multi-leader-follower
  stackelberg game using an evolutionary algorithm.
\newblock {\em Computers \& Operations Research}, 41:374--385.

\bibitem[Sinha et~al., 2006]{my-cec06}
Sinha, A., Srinivasan, A., and Deb, K. (2006).
\newblock A population-based, parent centric procedure for constrained
  real-parameter optimization.
\newblock In {\em 2006 IEEE Congress on Evolutionary Computation (CEC-2006)},
  pages 239--245. IEEE Press.

\bibitem[Sinha et~al., 2005]{my-cec05}
Sinha, A., Tiwari, S., and Deb, K. (2005).
\newblock A population-based, steady-state procedure for real-parameter
  optimization.
\newblock In {\em 2005 IEEE Congress on Evolutionary Computation (CEC-2005)},
  pages 514--521. IEEE Press.

\bibitem[Smith and Missen, 1982]{smith82}
Smith, W. and Missen, R. (1982).
\newblock {\em Chemical Reaction Equilibrium Analysis: Theory and Algorithms}.
\newblock John Wiley \& Sons, New York.

\bibitem[Sun et~al., 2008]{sun08}
Sun, H., Gao, Z., and Wu, J. (2008).
\newblock A bi-level programming model and solution algorithm for the location
  of logistics distribution centers.
\newblock {\em Applied Mathematical Modelling}, 32(4):610 -- 616.

\bibitem[Vicente and Calamai, 2004]{vicente-review}
Vicente, L.~N. and Calamai, P.~H. (2004).
\newblock Bilevel and multilevel programming: {A} bibliography review.
\newblock {\em Journal of Global Optimization}, 5(3):291--306.

\bibitem[Wang and Periaux, 2001]{stackelbergWang01}
Wang, F.~J. and Periaux, J. (2001).
\newblock Multi-point optimization using gas and {Nash/Stackelberg} games for
  high lift multi-airfoil design in aerodynamics.
\newblock In {\em Proceedings of the 2001 Congress on Evolutionary Computation
  (CEC-2001)}, pages 552--559.

\bibitem[Wang et~al., 2008]{GA_Wang}
Wang, G., Wan, Z., Wang, X., and Lv, Y. (2008).
\newblock Genetic algorithm based on simplex method for solving
  linear-quadratic bilevel programming problem.
\newblock {\em Comput Math Appl}, 56(10):2550--2555.

\bibitem[Yin, 2000]{yin-bilevel}
Yin, Y. (2000).
\newblock Genetic algorithm based approach for bilevel programming models.
\newblock {\em Journal of Transportation Engineering}, 126(2):115--120.

\end{thebibliography}

\end{document}